\documentclass[letterpaper,11pt]{article}
\pdfoutput=1
\usepackage{jheppub}
\usepackage{slashed}
\usepackage{graphicx}
\usepackage{subfigure}
\usepackage{soul}
\usepackage{amsmath}
\usepackage{multirow}
\usepackage{mathtools}
\usepackage{hyperref}
\usepackage{epstopdf}

\newcommand{\beq}{\begin{equation}}
\newcommand{\eeq}{\end{equation}}
\newcommand{\bea}{\begin{eqnarray}}
\newcommand{\eea}{\end{eqnarray}}

\def\m1{M_1}
\def\m2{M_2}
\def\m3{M_3}

\def\ch10{\tilde \chi^0_1}

\def\BR{\rm Br}

\def\gev{\,{\rm GeV}}

\def\to{\rightarrow}

\newcommand{\lsim}{\mathrel{\mathop{\kern 0pt \rlap
  {\raise.2ex\hbox{$<$}}}
  \lower.9ex\hbox{\kern-.190em $\sim$}}}
\newcommand{\gsim}{\mathrel{\mathop{\kern 0pt \rlap
  {\raise.2ex\hbox{$>$}}}
  \lower.9ex\hbox{\kern-.190em $\sim$}}}

\definecolor{pink}{RGB}{255,105,180}

\newcommand{\fb}{{\,{\rm fb}}}

\title{Running after Diphoton}

\author[a]{Jiayin Gu\footnote{Email: gujy@ihep.ac.cn}}
\author[b]{, Zhen Liu\footnote{Email: zliu2@fnal.gov}}

\affiliation[a]{Center for Future High Energy Physics, Institute of High Energy Physics, \\ Chinese Academy of Sciences, Beijing 100049, China}
\affiliation[b]{Theoretical Physics Department, Fermi National Accelerator Laboratory, Batavia, IL, 60510}

\abstract{
A very plausible explanation for the recently observed diphoton excess at the 13\,TeV LHC is a (pseudo)scalar with mass around $750$\,GeV, which couples to a gluon pair and to a photon pair through loops involving vector-like quarks (VLQs).  To accommodate the observed rate, the required Yukawa couplings tend to be large. A large Yukawa coupling would rapidly run up with the scale and quickly reach the perturbativity bound, indicating that new physics, possibly with a strong dynamics origin, is near by.  The case becomes stronger especially if the ATLAS observation of a large width persists. In this paper we study the implication on the scale of new physics from the 750 GeV diphoton excess using the method of renormalization group running with careful treatment of different contributions and perturbativity criterion.  
Our results suggest that the scale of new physics is generically not much larger than the TeV scale, in particular if the width of the hinted (pseudo)scalar is large.  Introducing multiple copies of VLQs, lowing the VLQ masses and enlarging VLQ electric charges help reduce the required Yukawa couplings and can push the cutoff scale to higher values.  Nevertheless, if the width of the $750$\,GeV resonance turns out to be larger than about 1\,GeV, it is very hard to increase the cutoff scale beyond a few TeVs. This is a strong hint that new particles in addition to the $750$\,GeV resonance and the vector-like quarks should be around the TeV scale.
}

\preprint{
\begin{flushright}
FERMILAB-PUB-15-573-T
\end{flushright}
}

\begin{document}

\maketitle
\flushbottom

%%%%%%%%%%%%%%%%%%%%%%%%%%%%%%%%%%%%%%%%%%%%%%%%%%%%%%%%%
\section{Introduction}

The ATLAS and CMS collaborations have recently announced the observation of an excess in the diphoton channel at around $750$\,GeV with the first crop of data from 13\,TeV LHC
\cite{CMS-PAS-EXO-15-004, ATLAS-CONF-2015-081}.  Many have speculated it to be the first hint of physics Beyond the Standard Model (BSM).  A huge amount of efforts from the theory community has been put on possible explanations and implications of this excess~\cite{Harigaya:2015ezk, Mambrini:2015wyu, Backovic:2015fnp, Nakai:2015ptz, Buttazzo:2015txu, Franceschini:2015kwy, DiChiara:2015vdm, Angelescu:2015uiz, Knapen:2015dap, Pilaftsis:2015ycr, Ellis:2015oso, Bellazzini:2015nxw, Gupta:2015zzs, Molinaro:2015cwg, Higaki:2015jag, McDermott:2015sck, Low:2015qep, Petersson:2015mkr, Dutta:2015wqh, Cao:2015pto, Matsuzaki:2015che, Kobakhidze:2015ldh, Cox:2015ckc, Ahmed:2015uqt, Agrawal:2015dbf, Martinez:2015kmn, Becirevic:2015fmu, No:2015bsn, Demidov:2015zqn, Chao:2015ttq, Fichet:2015vvy, Curtin:2015jcv, Bian:2015kjt, Chakrabortty:2015hff, Csaki:2015vek, Falkowski:2015swt, Aloni:2015mxa, Bai:2015nbs, Gabrielli:2015dhk, Benbrik:2015fyz, Kim:2015ron, Alves:2015jgx, Megias:2015ory, Carpenter:2015ucu, Bernon:2015abk, Chao:2015nsm, Arun:2015ubr, Han:2015cty, Chang:2015bzc, Chakraborty:2015jvs, Ding:2015rxx, Han:2015dlp, Luo:2015yio, Chang:2015sdy, Bardhan:2015hcr, Feng:2015wil, Antipin:2015kgh, Wang:2015kuj, Cao:2015twy, Huang:2015evq, Liao:2015tow,Heckman:2015kqk,Dhuria:2015ufo,Bi:2015uqd, Kim:2015ksf, Berthier:2015vbb, Cho:2015nxy,Cline:2015msi,Bauer:2015boy,Chala:2015cev,Kulkarni:2015gzu,Barducci:2015gtd, Boucenna:2015pav, Murphy:2015kag, Hernandez:2015ywg, Dey:2015bur, Pelaggi:2015knk, deBlas:2015hlv, Belyaev:2015hgo, Dev:2015isx, Huang:2015rkj, Hamada:2015skp}.
The majority of these studies involve a scalar or a pseudoscalar (usually denoted as $S$) with mass around $750$\,GeV, which couples to gluon pair and to photon pair through loops with vector-like quarks (VLQ), despite a few exceptions.
These explanations require large coefficients $c_g$ and $c_\gamma$ for the anomalous $Sgg$ and $S\gamma\gamma$ couplings.  Such large coefficients usually in turns require large Yukawa coupling between $S$ and the VLQs. In many cases this large coupling makes the theory already non-perturbative, 
indicating the invalidity of the assumed underlying BSM theory.  Even if the Yukawa coupling is below the perturbativity bound at $750$\,GeV, a sufficiently large Yukawa coupling will run up rapidly with scale. In this case the cutoff scale $\Lambda$ of the minimal theory, around which additional new physics needs to appear, can not be much larger than a few TeV.

In this paper, we study the implication of the scale of new physics from the 750 GeV diphoton excess by performing Rormalization Group (RG) running of the Yukawa couplings.  We explore different methods to increase the cutoff scale (thereby ``postpone'' the appearance of new physics). These methods include introducing multiple copies of VLQs, VLQ with light masses, and VLQ with large electric charges.  In addition, we consider different assumptions on the total decay width of the resonance, especially a relatively large width as preferred by the ATLAS results.  In Section~\ref{sec:basic}, we study the diphoton excess assuming the process $gg\to S \to \gamma\gamma$ account for all observed diphoton rate and derive the bounds on the couplings $c_g$ and $c_\gamma$ for different assumptions on the properties of the new (pseudo)scalar $S$ and the VLQs.  In Section~\ref{sec:RG}, we write down the RG equation of the Yukawa coupling and derive the cutoff scale with analytic method.  In Section~\ref{sec:num}, we perform a more careful numerical study and present the results for different cases.  We conclude in Section~\ref{sec:con}.

%%%%%%%%%%%%%%%%%%%%%%%%%%%%%%%%%%%%%%%%%%%%%%%%%%%%%%%%%

\section{Diphoton excess from $gg\to S \to \gamma\gamma$}
\label{sec:basic}
The observed cross section by combining both experiments at 8 TeV and 13 TeV under the assumption of $gg$-fusion production for a resonance $S$ is estimated to be \cite{Buttazzo:2015txu,Kim:2015ksf,Ellis:2015oso} 
\beq
\sigma(pp\to S)\times \BR(S\to\gamma\gamma) = 5.5\pm 1.4~fb~~{\rm at~LHC~13\,TeV}.
\eeq
We note here the acceptance times efficiency of the experiment is estimated at around 80\% from our simulation and is used in obtaining the above result. Our estimation of the acceptance times signal efficiency is consistent with the ATLAS description~\cite{ATLAS-CONF-2015-081}. 
The simplest model for such diphoton excess is a singlet scalar couples to gluon pair and photon pair through loop-induced anomalous couplings\footnote{ The phenomenology of such effective models has also been discussed in different context before the observation of diphoton excess, see {\it e.g.}~Ref.~\cite{Jaeckel:2012yz}. }. We consider the following effective Lagrangian in the broken phase of electroweak symmetry of a scalar (pseudoscalar) $s~(a)$ \footnote{Whenever possible, we use $S$ to simplify discussions applicable for both scalar $s$ and pseudoscalar $a$, as well as the case without ambiguities.}:
\bea
\mathcal{L}^s &&\supset -c_g \frac{\alpha_s}{12\pi} \frac s {m_S} G_{\mu\nu} G^{\mu\nu} - c_\gamma \frac {\alpha} {6\pi} \frac s {m_S} F^{\mu\nu} F_{\mu\nu}, \nonumber \\  
\mathcal{L}^a &&\supset \tilde c_g \frac{\alpha_s}{8\pi} \frac a {m_S} \epsilon_{\mu\nu\alpha\beta} G^{\mu\nu} G^{\alpha\beta} + \tilde c_\gamma \frac {\alpha} {4\pi} \frac a {m_S}  \epsilon_{\mu\nu\alpha\beta}F^{\mu\nu} F^{\alpha\beta}.
\label{eq:lagrangian}
\eea
We assume these couplings are generated by VLQ loops, which is the case in many plausible scenarios.
The normalization of these coefficients in Eq.~(\ref{eq:lagrangian}) are chosen such that their values basically correspond to the Yukawa couplings for a VLQ with its mass at $m_S=750~\gev$ for convenience.  To be more specific, we have
\bea
c_g,~\tilde c_g &&= \sum_i y_i \frac{m_S}{M_i} \bar A_{1/2}(\tau_i) \,,  \label{eq:cg}\\
c_\gamma,~\tilde c_\gamma &&= \sum_i y_i N_c Q_i^2\frac{m_S}{M_i} \bar A_{1/2}(\tau_i)  \,,  \label{eq:cga}
\eea
where $y_i$, $M_i$, $N_c$ and $Q_i$ are the corresponding Yukawa coupling, mass, number of colored states and electric charge of a given VLQ state.\footnote{We ignore any possible mixing between the VLQs and the SM quarks for simplicity. This mixing induces a coupling between the 750\,GeV resonance and SM quarks. This coupling introduces multiple mass scales and is constrained by other searches.  There also exist nontrivial bounds on the mixing between VLQs and the SM quarks, see {\it e.g.}~Ref.~\cite{Aguilar-Saavedra:2013qpa} for a comprehensive study.} Function $\bar A_{1/2}(\tau_i)$ is the correction to the loop-function of the given VLQ, with $\tau_i\equiv  {m_S^2}/( {4 M_i^2})$ and for $\tau_i < 1$~\cite{Gunion:1989we}
\bea
\bar A_{1/2}(\tau_i) =\begin{dcases}
    \frac 3 {2\tau_i^2}\left(\tau_i + (\tau_i - 1)\arcsin^2\sqrt{\tau_i}\right)\approx 1 + \frac 7 {30\tau_i}+O(\tau_i^2),& \text{for $s$}\\
    \frac 1 {\tau_i}\arcsin^2\sqrt{\tau_i}\approx 1+ \frac 1 {3\tau_i} +O(\tau_i^2) ,              & \text{for $a$}.
\end{dcases}
\label{eq:loopA}
\eea

\begin{figure}[t]
\centering
\includegraphics[width=10.0cm]{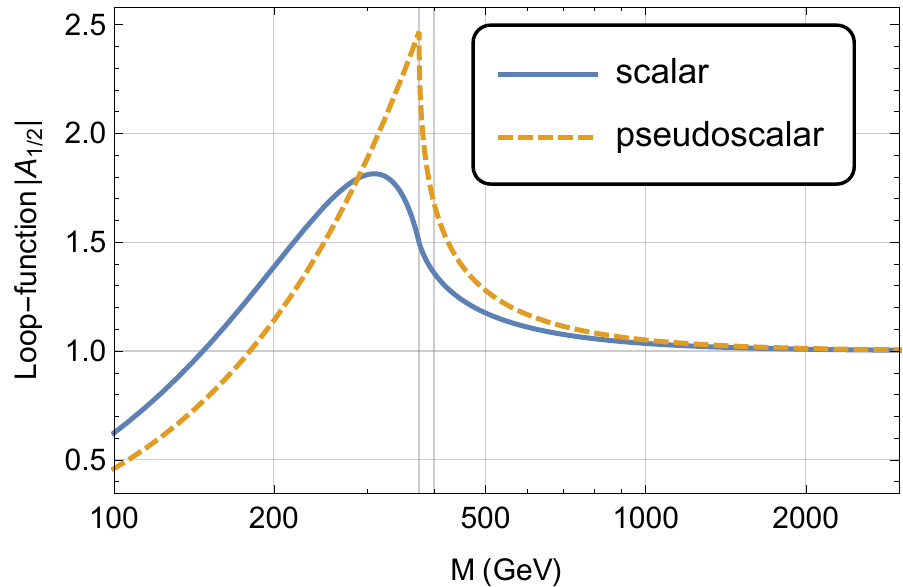}
\caption{The absolute value of the fermion Loop-function corrections for scalar and pseduoscalar as defined in Eq.(\ref{eq:loopA}).}
\label{fig:loop}
\end{figure}

In Fig.~\ref{fig:loop} we show the absolute values of these loop-function corrections as a function of the VLQ mass $M$ for a 750 GeV scalar particle. This demonstrates the size of near-threshold enhancement for the loop induced couplings. Comparing to the value of unity\footnote{This is due to our consistent choice of the coefficients in Eq.(\ref{eq:lagrangian}), Eq.(\ref{eq:cg}) and Eq.(\ref{eq:cga}).} in large $M$, the loop function could be enlarged to $1.5~(2.5)$ and 1.4~(1.7)  for VLQ mass $M$ 375\,GeV and 400\,GeV, respectively,  for scalar (pseudoscalar).  As we shall see later, the signal requires large values of loop induced scalar to gluon pair and to diphoton coupling. This near threshold effect is %important in relaxing the constraints of higher scale considerations. 
helpful in pushing up the cutoff scale of the theory.
This motivates our benchmark VLQ mass of 1000 GeV and 400 GeV. The former represents asymptotic values of loop function for large mass; the latter represents the case where threshold effect is important without opening up the tree-level two-body decays to VLQ pairs.

Before moving on to numerical studies of the diphoton excess, we want to comment on alternative choices of the effective Lagrangian. Conventionally we use the set of higher dimensional operators assuming the preservation of SM gauge symmetries $SU(3)_c\times SU(2)_L\times U(1)_Y$. It is not only a plausible requirement for most beyond stand model extensions, but also enables us to see the links between various modes after electroweak symmetry breaking. In such basis, the most relevant operators are $O_B$ and $O_W$\footnote {These operators are defined by replacing the $FF$ in operators in Eq.~(\ref{eq:lagrangian}) with $BB$ and $WW$ in the unbroken phase of electroweak gauge symmetry, and similarly for the pseudoscalar.}. Our coefficient $c_\gamma$ is in fact proportional to Wilson coefficient $c_B \cos^2\theta_W+c_W \sin^2\theta_W$. Furthermore, these operators also induce $S\to ZZ,~Z\gamma~{\rm and}~WW$ decays with explicit parametric dependence. Consequently, the future measurement of these relevant channels will provide new insight about the underlying theory, see discussions in {\it e.g.} Refs~\cite{Buttazzo:2015txu,Ellis:2015oso,Belyaev:2015hgo, Huo:2015exa}. However, our choice of parameterization does capture the essential physics for the diphoton anomaly, avoiding introducing more parameters to the question. Additional direct decay channels will also dilute the diphoton branching fractions, and in turn require larger coefficients. Our approach is rather minimalistic and conservative. A simple choice of $SU(2)_L$ singlet VLQ  will result in only inducing $O_B$ but not $O_W$. It was shown explicitly that the diphoton partial width is larger than the widths to other weak bosons in such setup~\cite{Low:2015qep, Bian:2015kjt, Altmannshofer:2015xfo, Low:2015qho}, following $\Gamma_{\gamma}:\Gamma_{Z\gamma}:\Gamma_{ZZ}=1: 2\tan\theta_W^2 : \tan\theta_W^4$ where $\theta_W$ is the Weinberg angle.  Other more complex models with VLQ charged under $SU(2)_L$ will lead to different relative strength of the $Z\gamma$, $ZZ$ and $WW$ partial widths, leading to interesting potential future searches to probe the underlying VLQ quantum numbers.

For general models with loop-induced scalar to gluon pair and photon pair coupling, one important feature is that the gluon pair partial width $\Gamma_{gg}$ is about $400$ times the diphoton partial width $\Gamma_{\gamma\gamma}$ for $c_g\approx c_\gamma$. We include in our calculation the important NLO K-factor for $\Gamma_{gg}$ of $1+67\alpha_s/(4\pi)$ from Ref.~\cite{Djouadi:2005gi} throughout this paper, which shifts the fitted result at O(1) level comparing to tree-level study. 
We first introduce the ``$\Gamma_{\rm min}$'' benchmark scenario when the scalar $S$ decay only contain the minimal set of partial widths $\Gamma_{\gamma\gamma}$ and $\Gamma_{gg}$ necessary for the observed diphoton excess, and additional partial widths of $\Gamma_{Z\gamma}$ and $\Gamma_{ZZ}$ together equals $0.7\,\Gamma_{\gamma\gamma}$ when $c_W=0$. This minimal width is at sub-GeV level for most parameter space we consider.  
However, the ATLAS observation of the diphoton excess at 13 TeV prefers a large width of around $45~\gev$. This motivates us to consider another two benchmark scenario of ``$\Gamma=1~\gev$'' as the medium width scenario and ``$\Gamma=45~\gev$'' as the large width scenario.

We obtain the production cross section for the scalar by scaling from the heavy SM Higgs production cross section reported by the Higgs cross section working group~\cite{Dittmaier:2011ti}, with appropriate luminosity ratios and loop function factors:
\beq
\sigma(gg\to S)=\frac {\sigma(gg\to H^{\rm SM}_{750\gev})} {\frac {m_H^2} {v^2} |\bar A(\frac {m_H^2} {4M_{\rm top}^2})|^2} \times c_g^2, \frac 9 4\tilde c_g^2= 55  c_g^2,~125 \tilde c_g^2~\fb,
\eeq
for $s$, $a$, respectively.  In the $\Gamma_{\rm min}$ scenario, the diphoton branching fraction is
\beq  
\BR(S\to\gamma\gamma)=\frac {c_\gamma^2} {1.7~c_\gamma^2 +  2 c_g^2 \frac{\alpha^2_s}{\alpha^2} (1+\frac {67} {4\pi} \alpha_s)},
\eeq
which is the same for the pseduoscalar.  For the benchmark scenario of $\Gamma=1\,\gev,~45\,\gev$, we require physical condition $\Gamma \ge\Gamma_{gg} +1.7~\Gamma_{\gamma\gamma}$. In above equation the factor of $1.7$ for the partial width  comes from the sum of all $\gamma\gamma~,Z\gamma$ and $ZZ$ partial widths in the $c_W=0$ case.  

\begin{figure}[t]
\centering
\includegraphics[width=7.5cm]{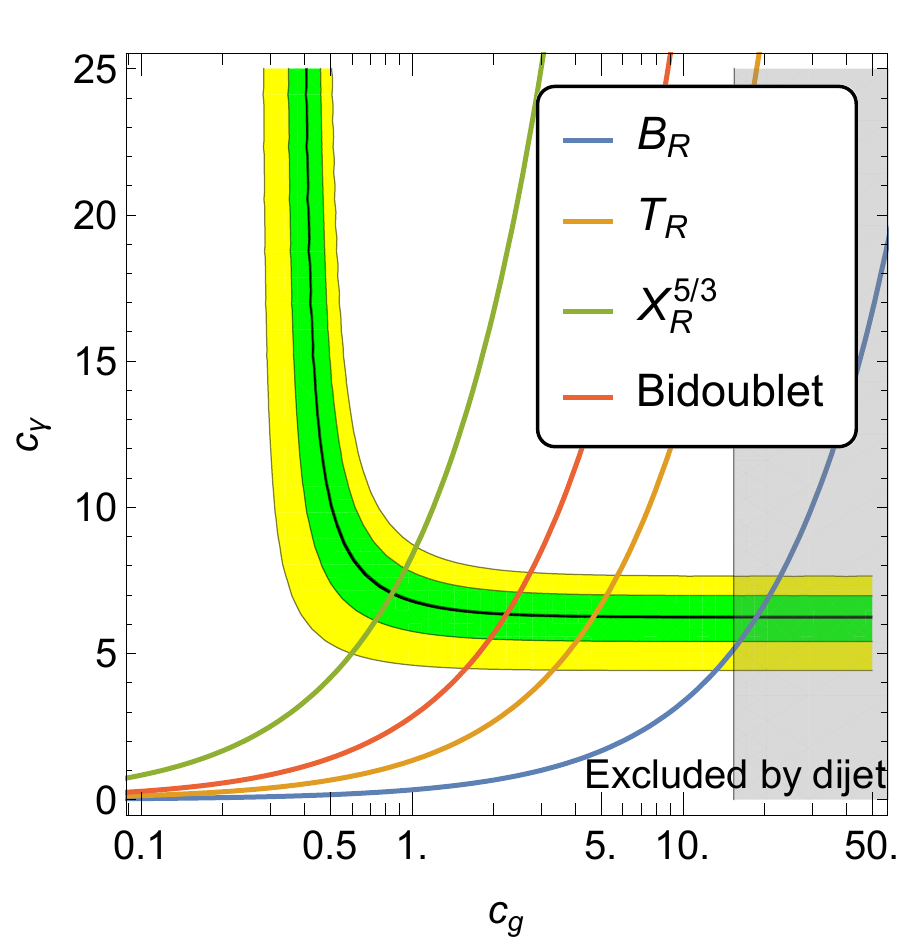}
\includegraphics[width=7.5cm]{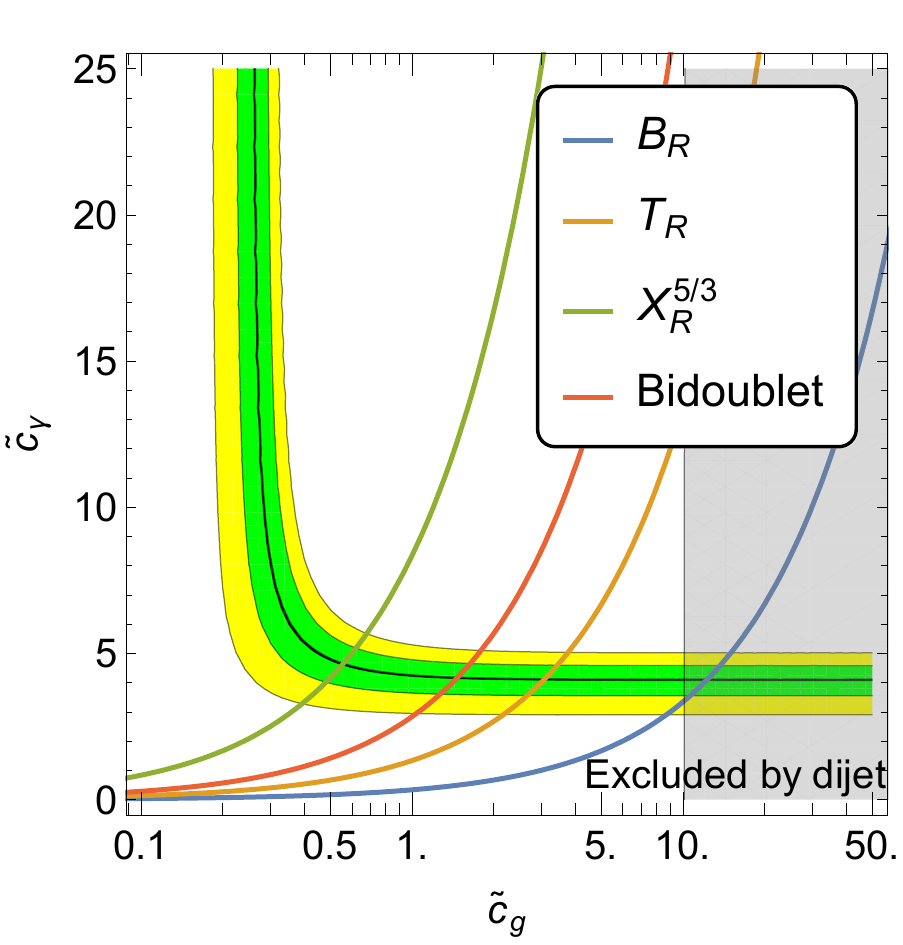}
\includegraphics[width=7.5cm]{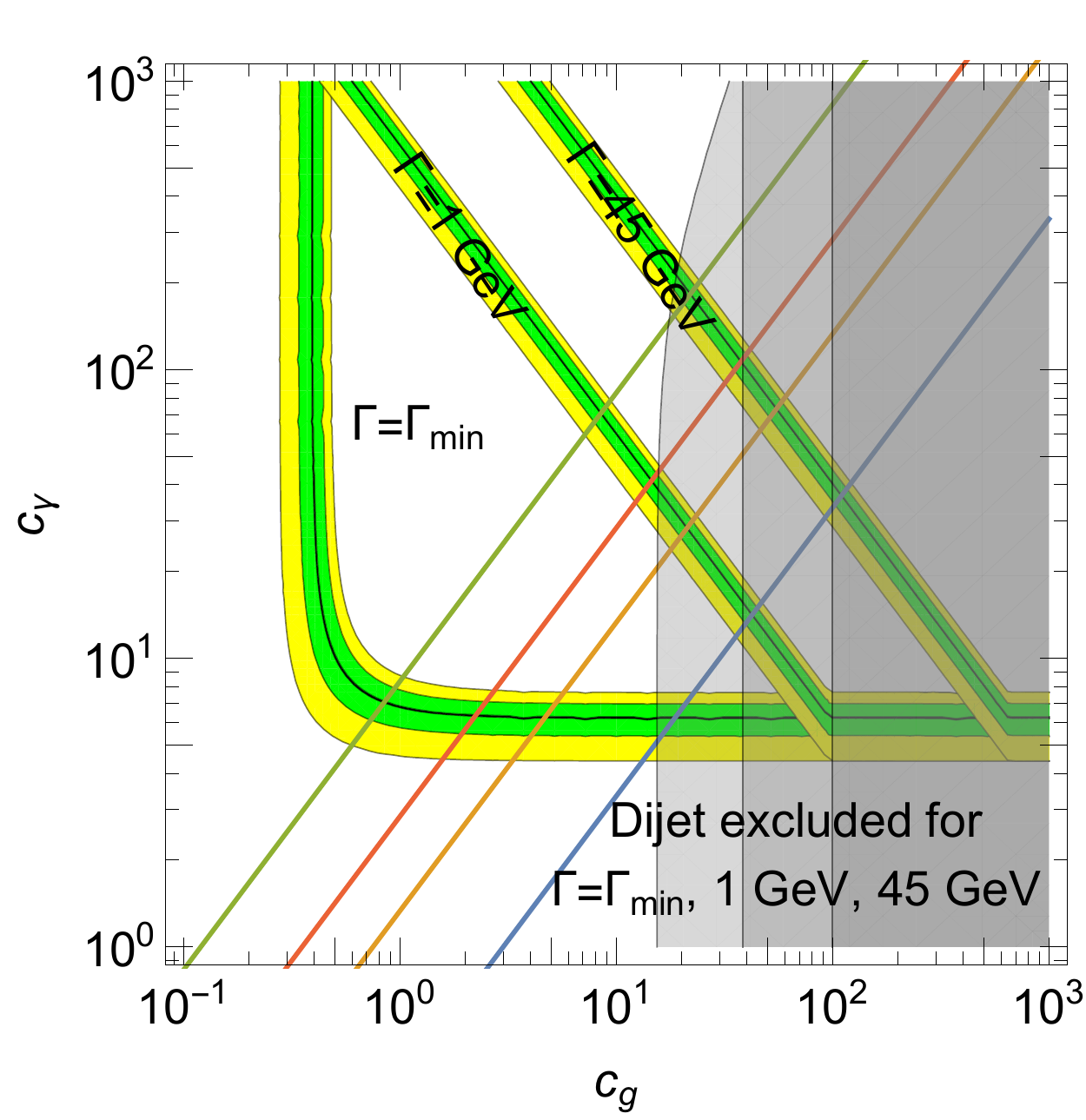}
\includegraphics[width=7.5cm]{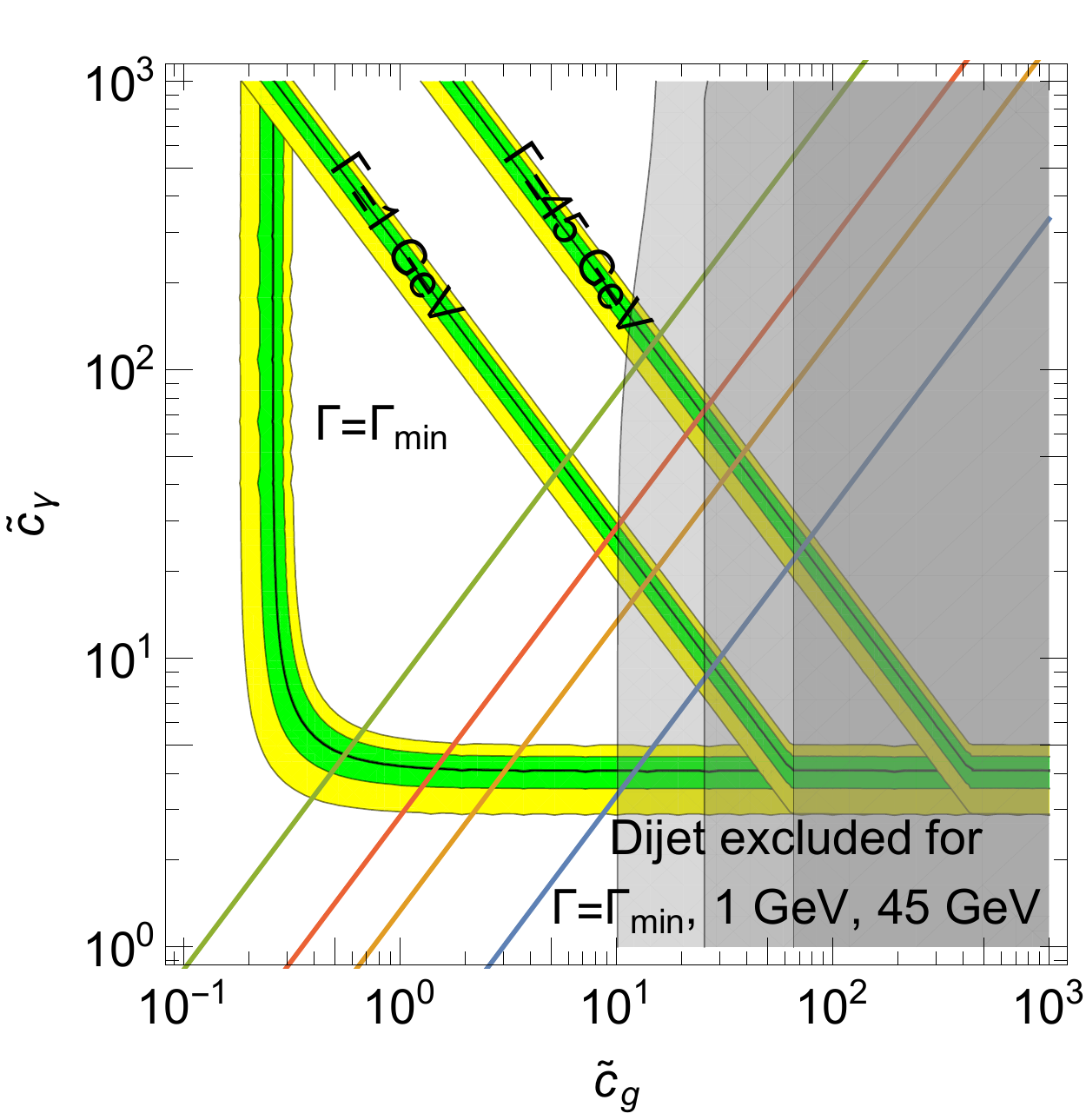}
\caption{Fitted values of $c_g$ and $c_\gamma$ with reference lines represent different VLQ typical models.  {\bf Top:} Results assuming $\Gamma=\Gamma_{\rm min}$. {\bf Bottom:} Results for $\Gamma=\Gamma_{\rm min}$, $\Gamma=1\,$GeV and $\Gamma=45\,$GeV are shown on the same plot.  The results for a scalar ($c_g$) is shown on the lefthand side and the ones for a pseudoscalar ($\tilde{c}_g$) is shown on the righthand side.  The green and yellow bands indicate the 68\% and 95\% confidence band of the fitted cross section, and the solid black line indicates the best fit value. The three different gray shaded region are parameter space excluded by the dijet search for $\Gamma=\Gamma_{\rm min}$, $\Gamma=1\,$GeV and $\Gamma=45\,$GeV, from light to dark, respectively.  Curves with different colors present the predicted relation for different VLQs ($B_R$, $T_R$, $X^{5/3}_R$ and the $(2,2)_{\frac 2 3}$ bidoublet).}
\label{fig:fit}
\end{figure}

The fitted values of the coefficients of $c_g$ and $c_\gamma$ ($\tilde c_g$ and $\tilde c_\gamma$) of the scalar (pseudoscalar) that accommodates the observed 750 GeV diphoton excess are shown in Fig.~\ref{fig:fit}.  The green and yellow bands indicate the 68\% and 95\% confidence band of the fitted cross section, and the solid black line indicates the best fit value.  
We show in the upper panel of this figure the $\Gamma_{\rm min}$ scenario and lower panel for all three scenarios together in logarithmic scale. Clearly, the vertical band for the $\Gamma_{\rm min}$ scenario corresponds to $\BR(S\to\gamma\gamma)$ dominance, where the value of $c_g$ controls the diphoton rate; the horizontal band for the $\Gamma_{\rm min}$ scenario corresponds to the $\BR(S\to gg )$ dominance, where the value of $c_\gamma$ controls the diphoton rate. For the scenarios with $\Gamma=1~\gev$ and 45~\gev, the diagonal bands corresponds to constant $c_g c_\gamma$ that controls the diphoton rate for a fixed total width.
We also apply an unavoidable dijet search cross section bound of $3\,$pb at the 8~TeV LHC~\cite{Aad:2014aqa,Khachatryan:2015sja}, shown in Fig.~\ref{fig:fit} with three different gray shaded regions for $\Gamma=\Gamma_{\rm min}$, $\Gamma=1\,$GeV and $\Gamma=45\,$GeV, from light to dark, respectively.  

\begin{table}[t]
\begin{center}
    \begin{tabular}{| c  | c | c |c|c|c|}
    \hline
    \multirow{3}{*}{Name} &   \multirow{3}{*}{$Q_{\rm em}$ states} &\multicolumn{4}{|c|}{$\sigma_{\gamma\gamma}$~Fits with various $\Gamma$}\\ \cline{3-6}
    & &   \multicolumn{2}{|c|}{$\Gamma_{\rm min}$, only $gg,~\gamma\gamma$} & 1 GeV & 45 GeV \\ \cline{3-6}
    & &   $c_g$ & $c_\gamma$ &  $c_g$ & $c_g$\\ \hline \hline
    $B_R$ &  $-\frac 1 3$ & $18.7^{+2.2}_{-2.5}$  & $6.24^{+0.74}_{-0.84}$ & $42.5^{+2.4}_{-2.9}$ & $110^{+6}_{-8}$\\ \hline
    $T_R$ &  $\frac 2 3$ & $4.70^{+0.55}_{-0.63}$ & $6.26^{+0.74}_{-0.84}$ & $21.2^{+1.2}_{-1.5}$ & $55.0^{+3.2}_{-3.8}$\\ \hline
    $X_R^{5/3}$ & $\frac 5 3$ & $0.85^{+0.10}_{-0.11}$ & $7.06^{+0.83}_{-0.95}$ & $8.50^{+0.49}_{-0.59}$ & $22.0^{+1.3}_{-1.5}$\\ \hline
    Bidoublet &   $-\frac 1 3~+\frac 2 3~+\frac 2 3~+\frac 5 3$ & $2.24^{+0.26}_{-0.30}$ & $6.34^{+0.75}_{-0.85}$ & $14.6^{+0.8}_{-1.0}$ & $37.7^{+2.2}_{-2.6}$\\ \hline
    \hline
    Name & $Q_{\rm em}$ states &   $\tilde c_g$ & $\tilde c_\gamma$ &  $\tilde c_g$ & $\tilde c_g$\\ \hline
$B_R$ &  $-\frac 1 3$ & $12.3^{+1.5}_{-1.6}$  & $4.09^{+0.48}_{-0.55}$ & $28.1^{+1.6}_{-2.0}$ & $72.8^{+4.2}_{-5.1}$\\ \hline
    $T_R$ &  $\frac 2 3$ & $3.08^{+0.36}_{-0.41}$ & $4.11^{+0.49}_{-0.55}$ & $14.1^{+0.8}_{-1.0}$ & $36.4^{+2.1}_{-2.5}$\\ \hline
    $X_R^{5/3}$ & $\frac 5 3$ & $0.56^{+0.07}_{-0.08}$ & $4.63^{+0.55}_{-0.62}$ & $5.62^{+0.32}_{-0.39}$ & $14.6^{+0.8}_{-1.0}$\\ \hline
    Bidoublet &   $-\frac 1 3~+\frac 2 3~+\frac 2 3~+\frac 5 3$ & $1.47^{+0.17}_{-0.20}$ & $4.16^{+0.49}_{-0.56}$ & $9.64^{+0.55}_{-0.67}$ & $25.0^{+1.4}_{-1.7}$\\ \hline
    \end{tabular}
\caption{The $1\sigma$ preferred range on the loop-induced coupling coefficient from the observed diphoton excess for different VLQ cases, which are $B_R$, $T_R$, $X^{5/3}_R$ and the $(2,2)_{\frac 2 3}$ bidoublet. Here we assume all VLQs are in the fundamental representation of $SU(3)_c$.  The results for a scalar (pseudoscalar) resonance are shown on the top (bottom) half of the table.   We also show the results for different assumptions on the decay width of the resonance.  As in each case $c_g$($\tilde{c}_g$) and $c_\gamma$($\tilde{c}_\gamma$) are not independent and for VLQs, it is more convenient to parameterize in terms of $c_g$($\tilde{c}_g$). For $\Gamma=1\,$GeV and 45\,GeV we only show the results for $c_g$($\tilde{c}_g$).
}
\label{tab:coefficients}
\end{center}
\end{table}

For a given VLQ with certain charge and representation, the value of $c_g$ and $c_\gamma$ ($\tilde{c}_g$ and $\tilde{c}_\gamma$) are not independent, as predicted by Eq.~(\ref{eq:cg}) and Eq.~(\ref{eq:cga}).  In Fig.~\ref{fig:fit} we also show the predicted relations between the two coefficients for VLQs with different charges and representations, each with a curve of a different color.  We have chosen a few representative cases: three singlet VLQs with electric charges $1/3$, $2/3$ and $5/3$ (denoted as $B_R$, $T_R$ and $X^{5/3}_R$, respectively, where the subscript ``$R$'' simply indicates them being SM $SU(2)_L$ singlets) and an $SU(2)_L\times SU(2)_R$ bidoublet with ${U(1)_X}$ hypercharge $2/3$ (often denoted as $(2,2)_{\frac 2 3}$), motivated by natural composite Higgs models~\cite{Contino:2006qr}.\footnote{The bidoublet induces none-zero $c_W$, leading to new decay channels of $S\to WW$ and a sum of all electroweak boson partial widths different  to the $1.7~\Gamma_{\gamma\gamma}$ of the $c_W=0$ case. For very small $c_g$ (vertical bands in Fig.~\ref{fig:fit}) this will shift the fitted $c_g$ to larger values. Still, as the line for the predicted ratio $c_\gamma/c_g$ intersects with the $S\to gg$ dominant region (horizontal bands) the fitted values will remain the same and thus we show it on a same plot.} These  cases represent various possibilities while keeping the analysis relatively simple.  It is straight forward to read the constraints on values of both $c_g$ and $c_\gamma$ ($\tilde{c}_g$ and $\tilde{c}_\gamma$) from the intersection of the curves and the green/yellow bands.  We present the $1\sigma$ preferred range on the required $c_g$ ($\tilde{c}_g$) in Table~\ref{tab:coefficients} for different VLQ cases.  These values are used later in Section~\ref{sec:num} for estimating the cutoff scale for different VLQs.

%%%%%%%%%%%%%%%%%%%%%%%%%%%%%%%%%%%%%%%%%%%%%%%%%%%%%%%%%%%%

\section{Perturbative considerations with RG running}
\label{sec:RG}

In this section we try to study the RG running of the Yukawa coupling.  As discussed in the previous section a large Yukawa coupling between the singlet scalar and VLQ is expected in all scenarios, especially for $S$ with large width. Such large coupling, even if perturbative at scale $\mu=m_S$, is likely to violate perturbativity at nearby scale. This leads to our consideration of the RG running of these couplings. The logic behind is to access the validity of using $S$ and VLQs to explain diphoton excess, and to provide a conservative estimation of the scale of new physics beyond this minimal setup. An important and intuitive modification in the case of large $c_g$ is to introduce multiple ($N_f$) copies of the VLQ, so as to reduce the required Yukawa coupling for each flavor.  We discuss in details the modification associated with such introduction of $N_f$ in both RG running and perturbativity criterion in the following two sections.

\begin{figure}[t]
\centering
\includegraphics[width=15cm]{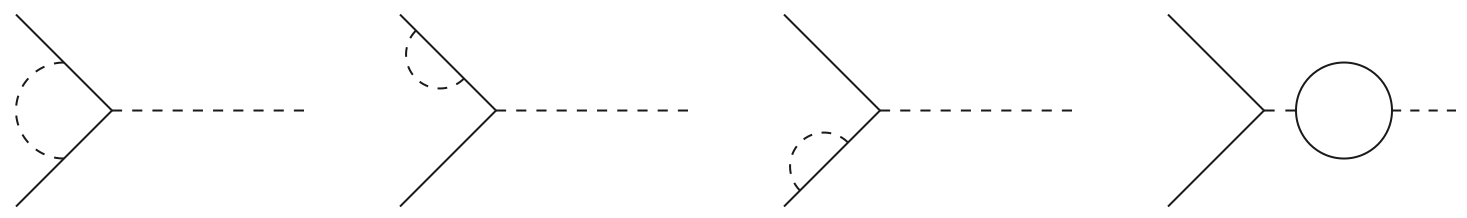}
\caption{Feynman diagrams for one loop corrections to the Yukawa coupling $y$ corresponding to the $y^3$ term in the RG equation of $y$ in Eq.~(\ref{eq:rgy}).  
The total contribution from the 4 diagrams is proportional to $(3 + 2 N_c N_f )\, y^3$, where the dominant contribution is from the last diagram in which the closed fermion loop gives a factor of $N_c N_f$.}
\label{fig:yukawa1}
\end{figure}

To simplify calculation while capturing most of the physics, we assume the scalar or pseudoscalar  
couples to $N_f$ copies of VLQs with the same masses $M$ and a diagonal Yukawa coupling matrix $y \, \boldsymbol{\rm I}$ where $ \boldsymbol{\rm I}$ is an $N_f \times N_f$ identity matrix.  Using the results from Ref.~\cite{Luo:2002ti}, we obtain the RG equation of the Yukawa coupling $y$, given by\footnote{Our current Eq.~(\ref{eq:rgy}) also agrees with the results in Ref.~\cite{Son:2015vfl, Goertz:2015nkp}. }
\begin{equation}
16\pi^2 \frac{d y}{dt} =(3 + 6 N_f )\, y^3 -8 g^2_3 \, y \, ,   \label{eq:rgy}
\end{equation}
where we have used $N_c=3$ for the number of colors. We have also ignored the contributions from electroweak gauge bosons, which is a good approximation for the range of VLQ charges ($|Q|\leq 5/3$) considered in this paper%\footnote{For a  $SU(2)_L$singlet VLQ with $|Q|= 5/3$, the contribution from electroweak gauge boson loops to Eq.~(\ref{eq:rgy}) is about $24\%$ of the one from the gluon loops.}
.  The first term of the righthand side of Eq.~(\ref{eq:rgy}) comes from scalar and fermion loops shown in Fig.~\ref{fig:yukawa1} and the second term is from gluon loops.  For a given scenario one could deduce the required values of $c_g$ and $c_\gamma$ (or $\tilde{c}_g$ and $\tilde{c}_\gamma$), which can be transformed into the desired value of $y$ at scale $\mu = 750\,$GeV. We will denote the value of $y$ at scale $\mu = 750\,$GeV as $y_0$ in order to distinguish it from the running value.  For convenience, we choose to parameterize $y_0$ in terms of the coupling to gluons, $c_g$~\footnote{The same derivation also works for a pseudoscalar with $\tilde{c}_g$.}.  Under our assumption, Eq~(\ref{eq:cg}) reduces to
\begin{equation}
c_g = A \, y_0 \, N_f \frac{m_S}{M} \,,  \label{eq:cgsimp}
\end{equation}
where we have denoted $\bar{A}_{1/2}(\frac{m^2_S}{4 M^2})$ simply as $A$.  Therefore, the initial value of the Yukawa coupling is given by
\begin{equation}
y_0 = \frac{c_g M}{A \, N_f m_S}  \,.  \label{eq:yre}
\end{equation}

We now proceed to evaluate the cutoff scale as a function of the desired value of $c_g$.  One could obtain an intuitive understanding of the running of $y$ by making a number of approximations.  First, we ignore the gluon contribution to the RG running which is small as long as $y$ is sufficiently large.  In this case, Eq.~(\ref{eq:rgy}) can be easily solved to give    
\begin{equation}
\frac{1}{y_0^2} - \frac{1}{y^2} = \frac{3+ 6 N_f}{8\pi^2} \log{(\frac{\mu}{\mu_0})} \,, \label{eq:resimp1}
\end{equation}
where $y_0$ is given by Eq.~(\ref{eq:yre}) and $\mu_0$ is the initial scale for the running at which $y=y_0$, given by 
\begin{equation}
\mu_0 = \left\{ \begin{matrix}  m_S \mbox{ ~ if  } M < m_S  \\ M  \mbox{  \, ~~if  } M \ge m_S \end{matrix}  \right. \,,
\end{equation}
as the running only starts above scale $M$.  To simplify calculation we also assume the cutoff scale $\Lambda$ is the scale at which $y\to \infty$.   A more appropriate definition of the cutoff scale should be the one at which the couplings become non-perturbative, which we will address in the next section.  From Eq.~(\ref{eq:resimp1}), we have
\begin{align}
\log{(\frac{\Lambda}{\mu_0})} =& ~ \frac{8\pi^2}{3+6 N_f} \left( \frac{A_f N_f m_S}{c_g M}  \right)^2   \nonumber\\  
\approx & ~ \frac{4\pi^2}{3} \frac{N_f}{c_g^2}  \left( \frac{A_f m_S}{ M}  \right)^2  \label{eq:cutoff2} \,,
\end{align}
where in the last line we assumed $6N_f \gg 3$.  The righthand side of Eq.~(\ref{eq:cutoff2}) is proportional to $N^2_f \cdot N^{-1}_f = N_f$, where $N^2_f$ comes from the fact that increasing $N_f$ reduces the initial Yukawa coupling $y_0$ needed to obtain a certain value of $c_g$, and $N^{-1}_f$ comes from the fact that $N_f$ enhances the RG evolution of $y$.  Therefore, a crucial result here is that $\log{\Lambda}$ is proportional to $N_f/c_g^2$.  A very large $N_f$ helps increase $\Lambda$, while a large $c_g$ would force $\Lambda$ to be small.

%%%%%%%%%%%%%%%%%%%%%%%%%%%%%%%%%%%%%%%%%%%%%%%%%%%%%%%%%%%%

\section{Implication of new physics scale in various VLQ models}
\label{sec:num}

We now move on to a more careful numerical study of the cutoff scale $\Lambda$ of the minimal theory with the $750\,$GeV resonance $S$ and VLQs, at which scale the couplings become non-perturbative. We will use a more appropriate definition of the cutoff scale and also include the gluon contribution to the RG evolution of the Yukawa coupling.  
One usually consider $\Lambda$ to be given by the unitarity or perturbativity bound, which is different in descriptions but comes from the same physics origin\footnote{For a discussion on possible separation between the scale of new physics and the scale of perturbativity violation, see Ref.~\cite{Aydemir:2012nz}.}.  One could think of a 4-fermion scattering process with an s-channel $S$, which has an amplitude proportional to $N_c N_f \,y^2$, where the factor of $N_c N_f$ comes from the sum of final state fermions.  One could also think of the same process with a self energy term of $S$ involving fermion loop proportional to $\frac{N_c N_f \,y^2}{(4\pi)^2}$.  In both picture, $\Lambda$ is found be the scale at which $y \sim \frac{4\pi}{\sqrt{N_c N_f}}$.  
Therefore, one would expect a large $N_f$ to be less helpful in terms of increasing the cutoff scale than one would naively expect. Nevertheless, a large $N_f$ still helps increase $\Lambda$ since $y_0$ scales as $1/N_f$ for a fixed value of $c_g$.

The gluon contribution to the RG evolution of $y$ is given by the 2nd term on the righthand side of Eq.~(\ref{eq:rgy}).  For large $N_f$, the required $y_0$ is small and the gluon contribution could significantly slow down the running of $y$ or even make it run down\footnote{For some special regions of the parameter space, these couplings could approach an approximate conformal fixed point, leading to possible interesting links to higher scale physics.}. However, unless $N_f > 10$, $g_3$ would decrease as the scale goes up and, depending on the value of $y_0$, $y$ could either keep on running down or turn around at some point.  This means that for small enough $c_g$ or large enough $N_f$, the theory could be weakly coupled up to very high scales and essentially does not require a cutoff.  For example, as we will show later, if $S$ has a minimal total decay width and couples to a charge $5/3$ quark, it is possible to have a weakly coupled theory while achieving the desired diphoton excess.

\begin{figure}[t]
\centering
\includegraphics[width=7.5cm]{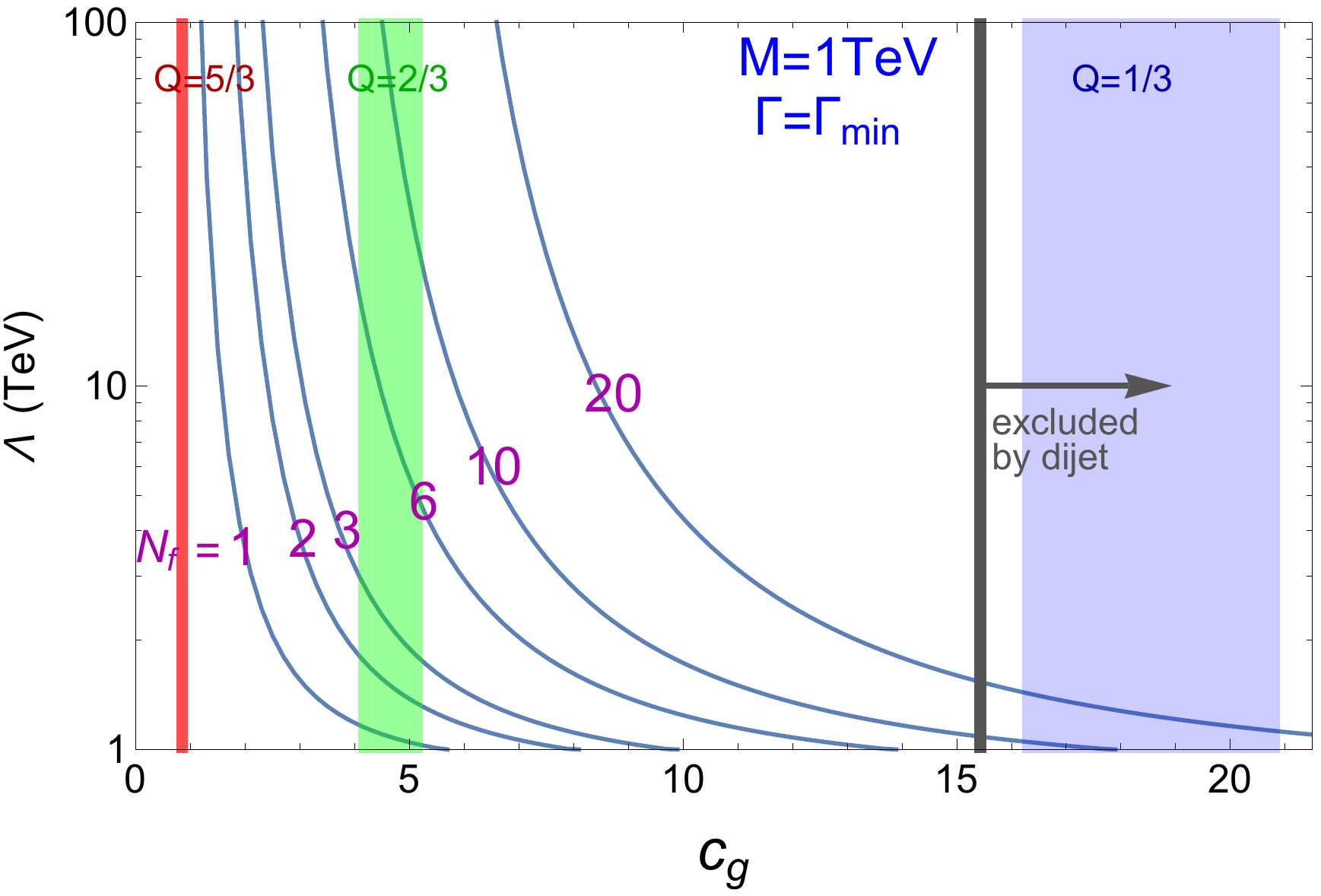}
\includegraphics[width=7.5cm]{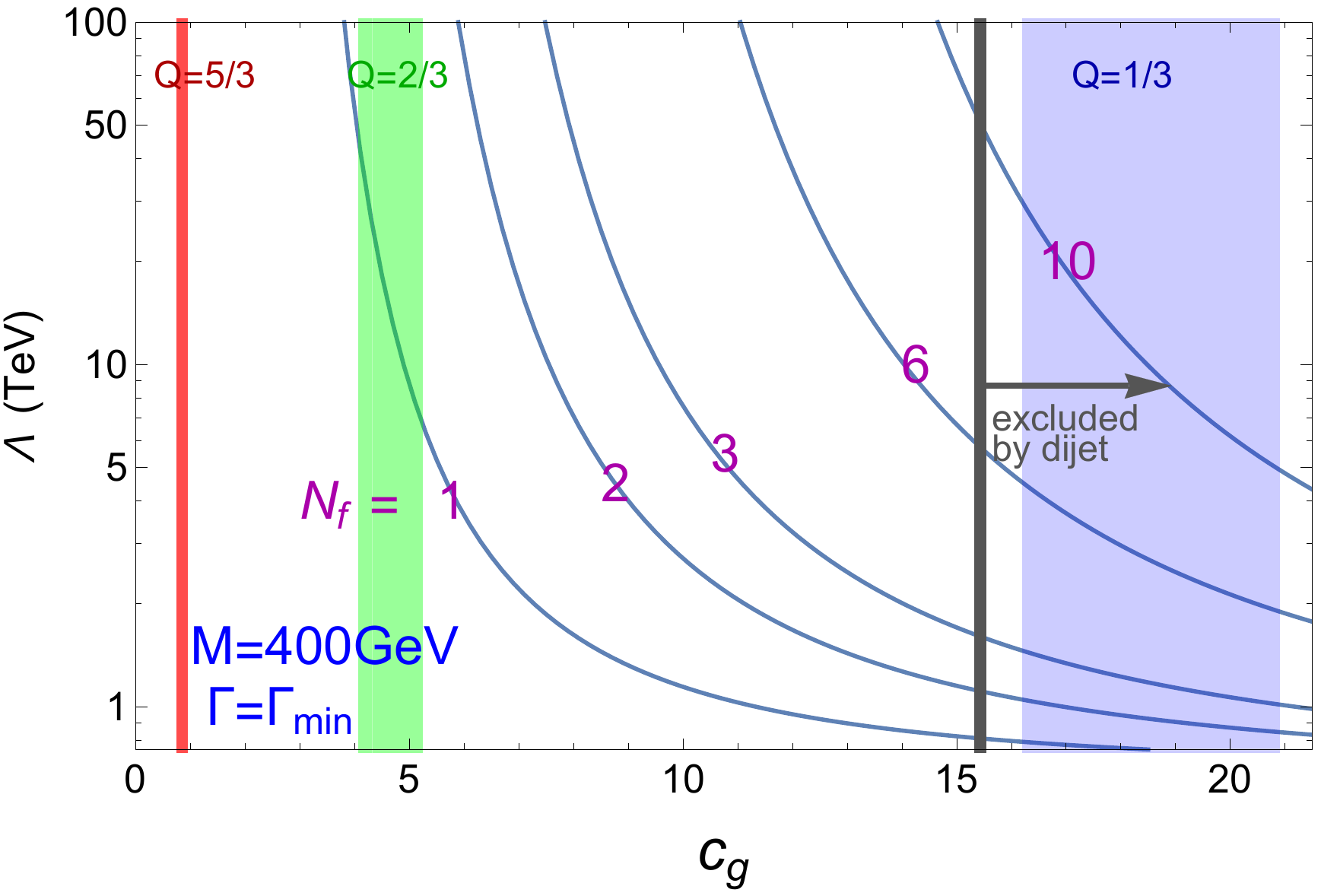} \\
\includegraphics[width=7.5cm]{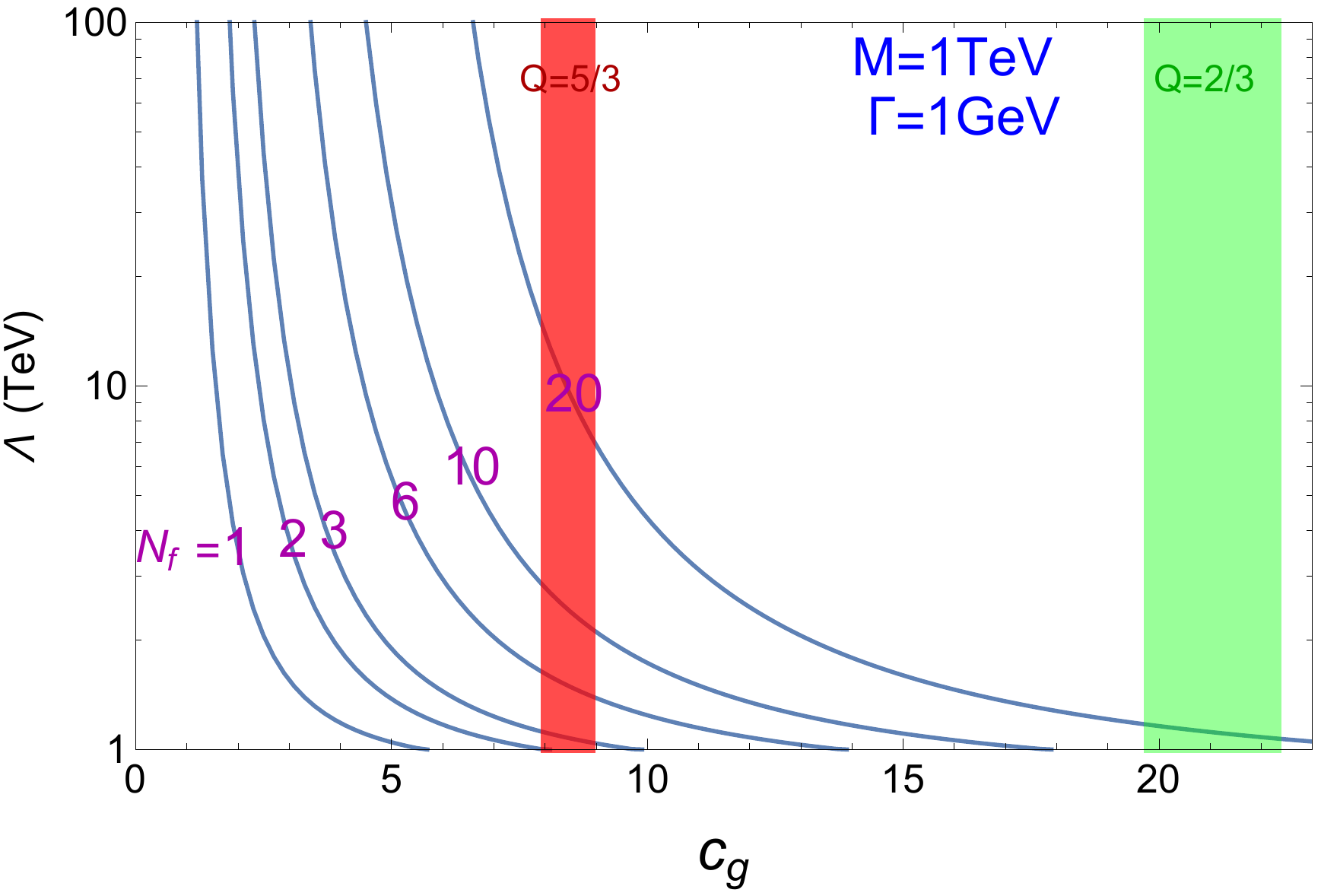}
\includegraphics[width=7.5cm]{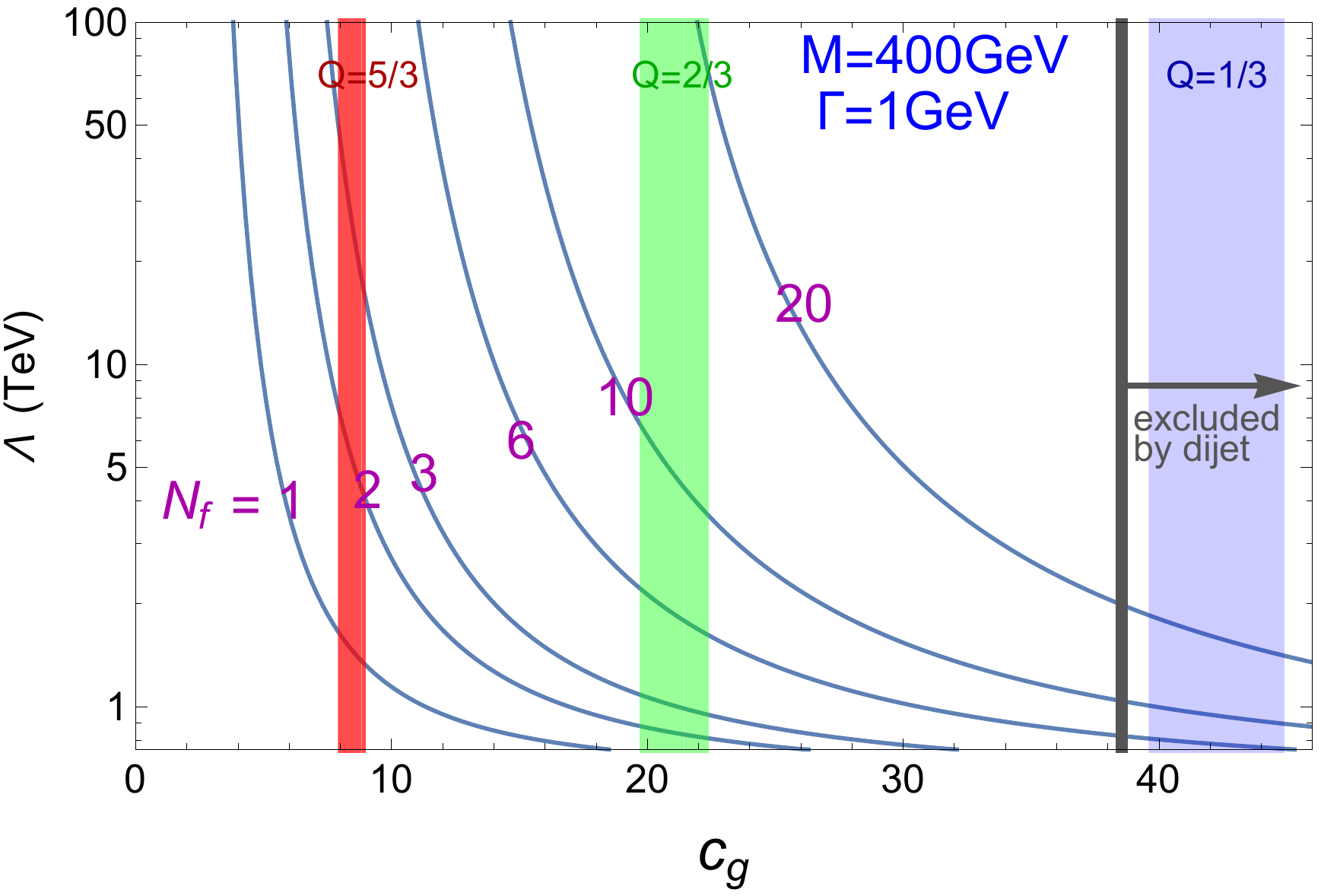} \\
\includegraphics[width=7.5cm]{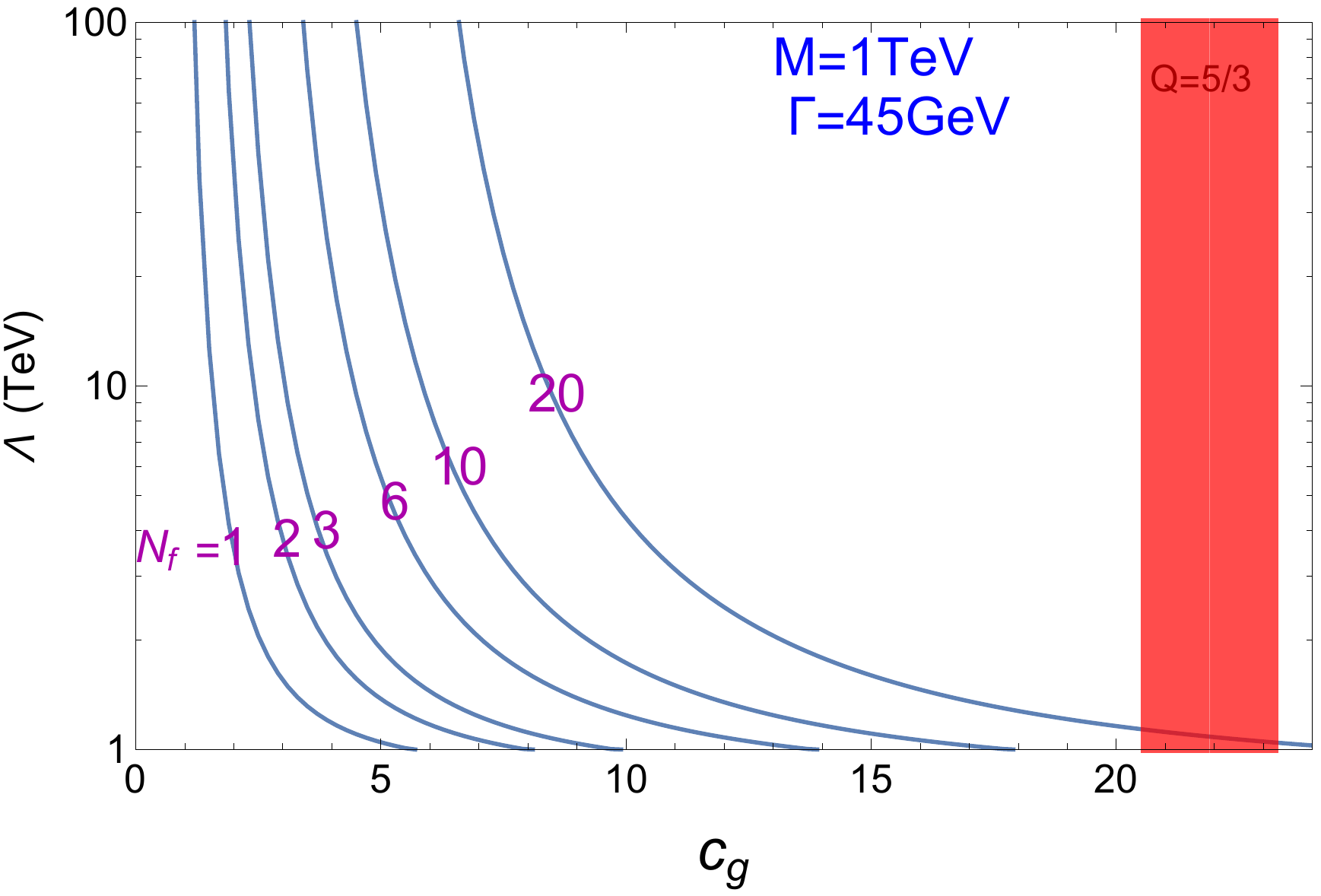}
\includegraphics[width=7.5cm]{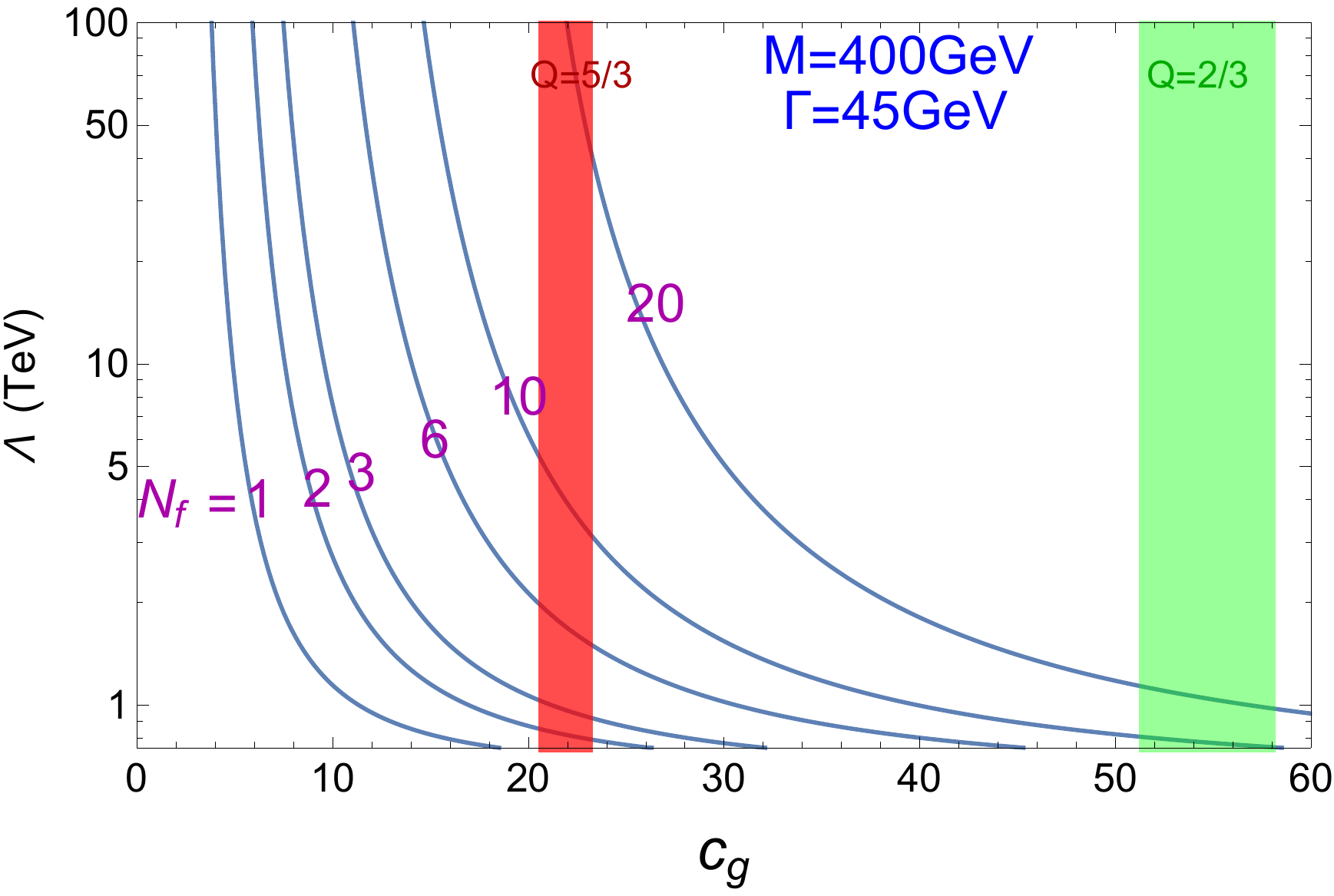}
\caption{The cutoff scale $\Lambda$ as a function of $c_g$ for different numbers of flavors $N_f$ with $m_S=750$\,GeV, for $S$ being a scalar.  {\bf Left:} $M=1$\,TeV,  {\bf Right:} $M=400$\,GeV.    
Different bands corresponds to the $1\sigma$ preferred ranges on $c_g$ required to produce the observed diphoton excess for singlet VLQs with different electric charges.  The red, green blue bands correspond to charge $5/3$, $2/3$ and $1/3$ quarks, respectively.  Different assumptions on the total width of $S$ are considered, which are $\Gamma = \Gamma_{\rm min}$ (top panel), $\Gamma =1\,$GeV (middle panel) and $\Gamma =45\,$GeV (bottom panel). The constraints on $c_g$ from the dijet searches are also indicated on the plots, which are $c_g < 15.4, ~38.5, ~100$ for $\Gamma = \Gamma_{\rm min}, ~1\,{\rm GeV},  ~45\,{\rm GeV}$, respectively.}
\label{fig:rge1}
\end{figure}
\begin{figure}[t]
\centering
\includegraphics[width=7.5cm]{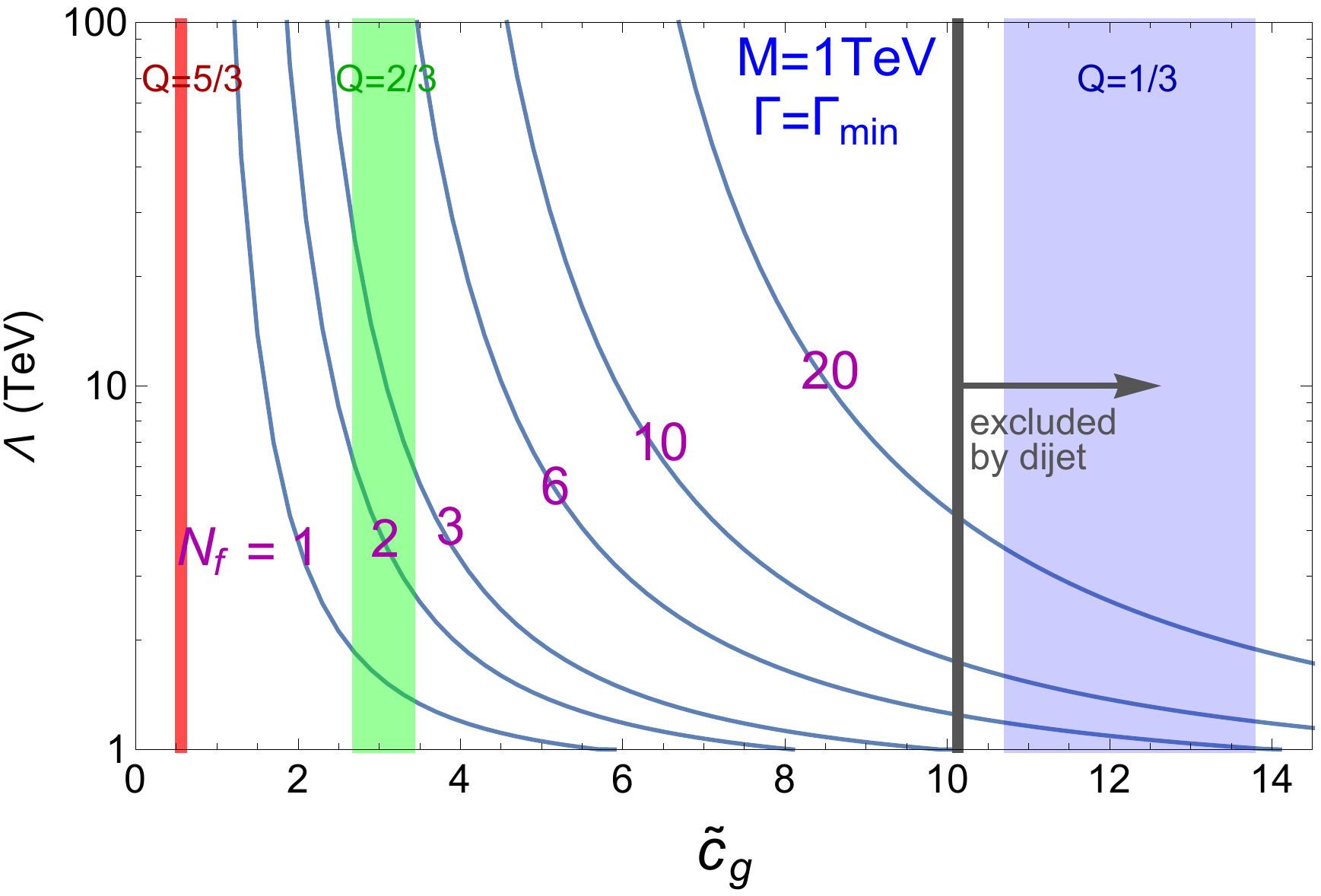}
\includegraphics[width=7.5cm]{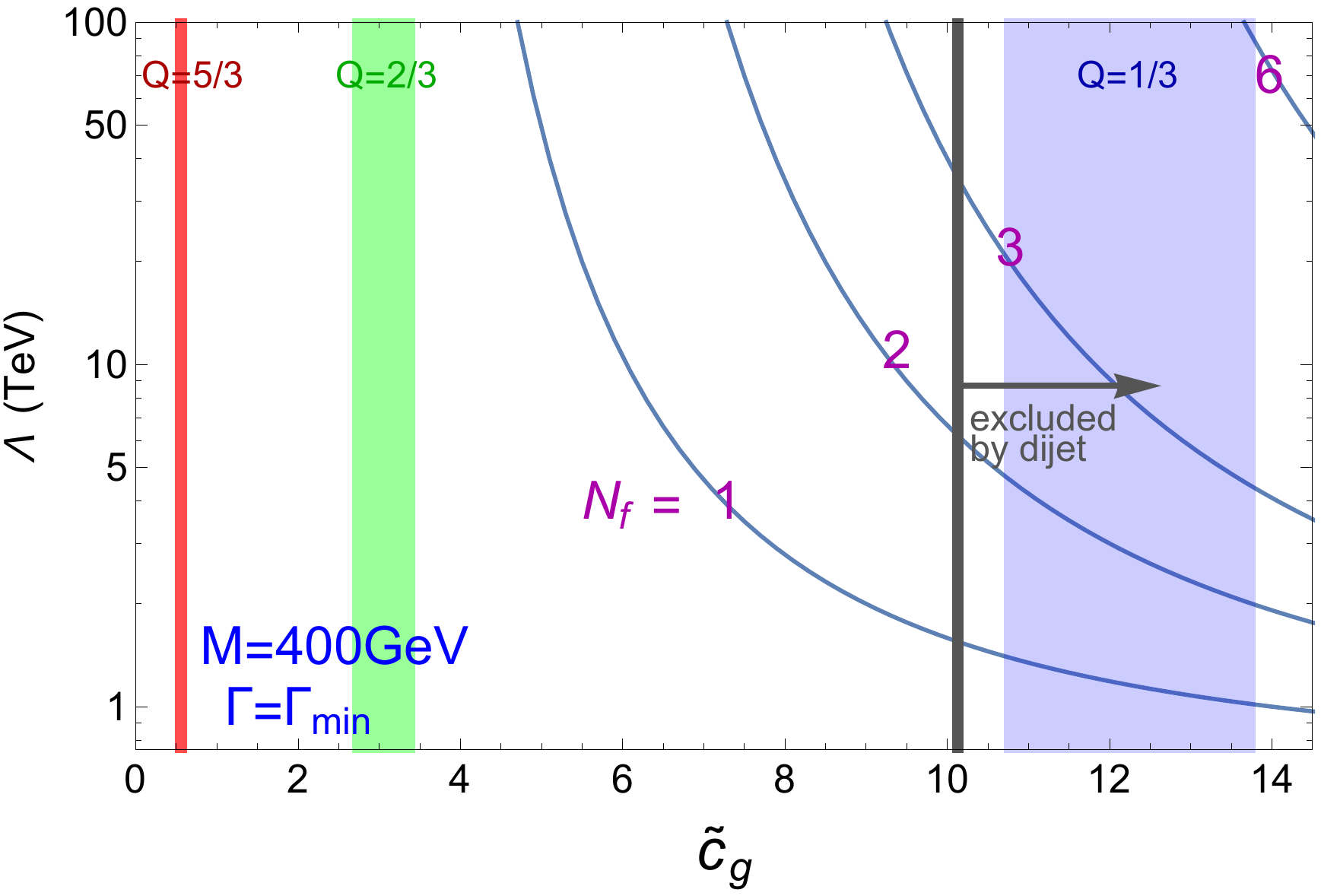} \\
\includegraphics[width=7.5cm]{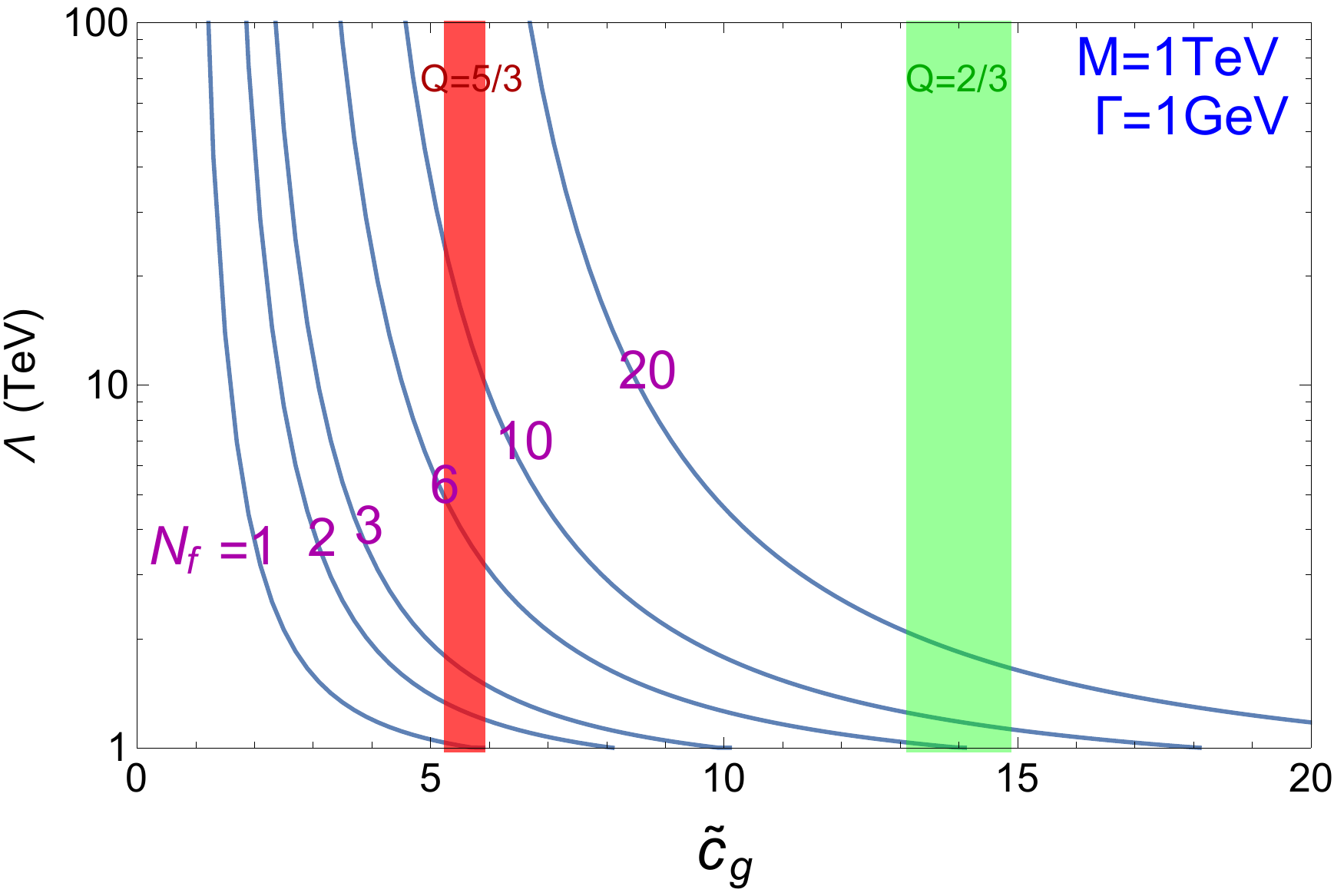}
\includegraphics[width=7.5cm]{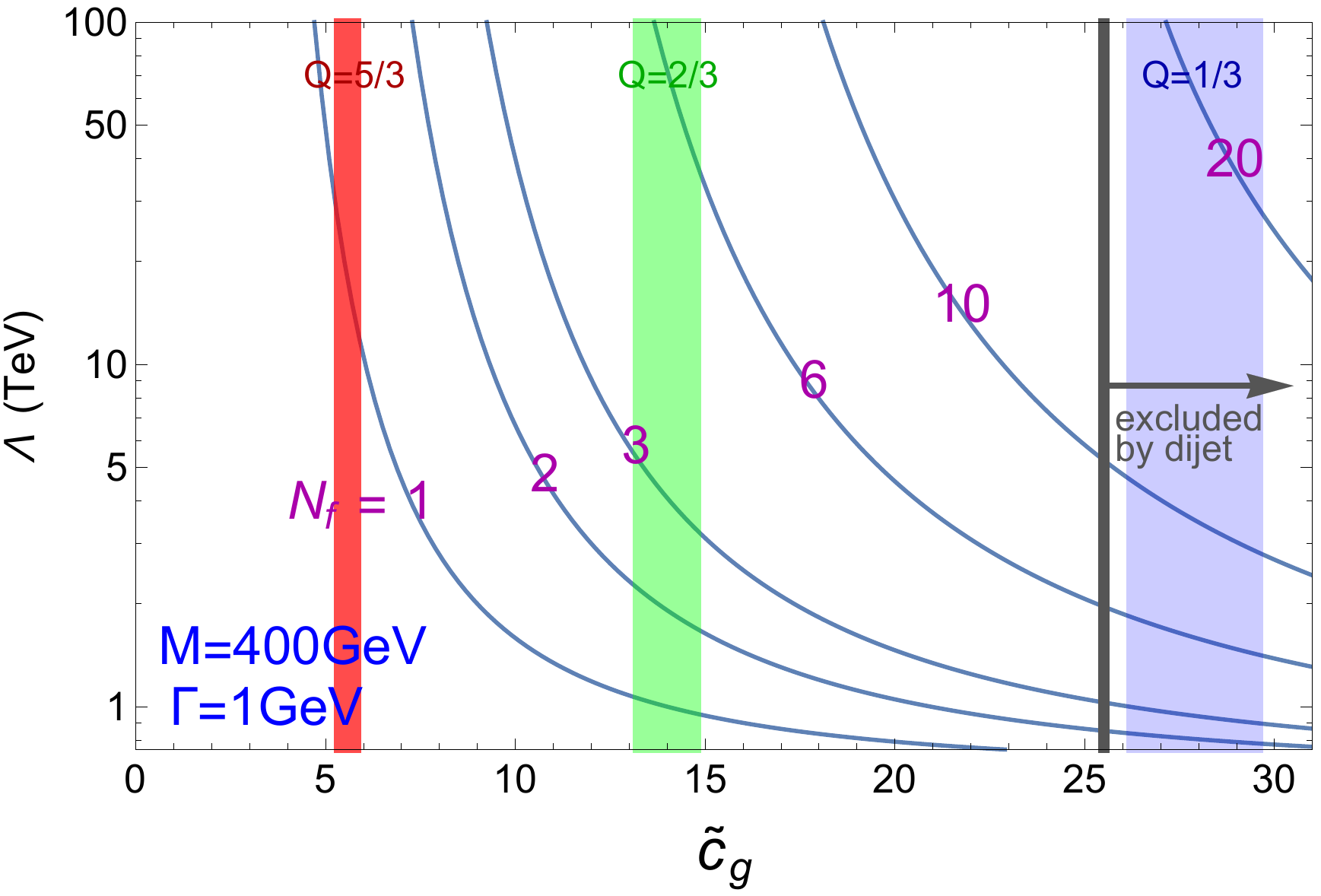} \\
\includegraphics[width=7.5cm]{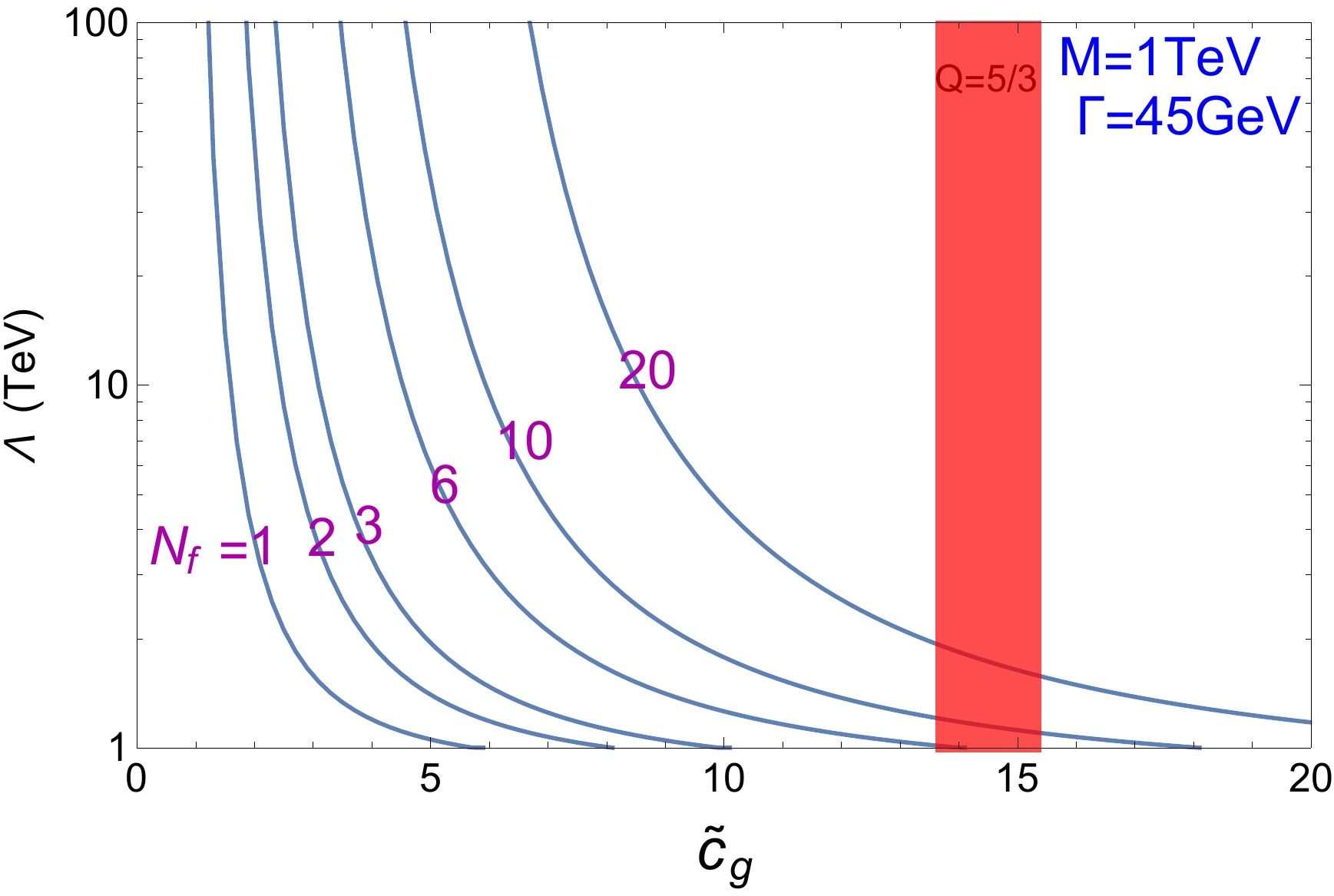}
\includegraphics[width=7.5cm]{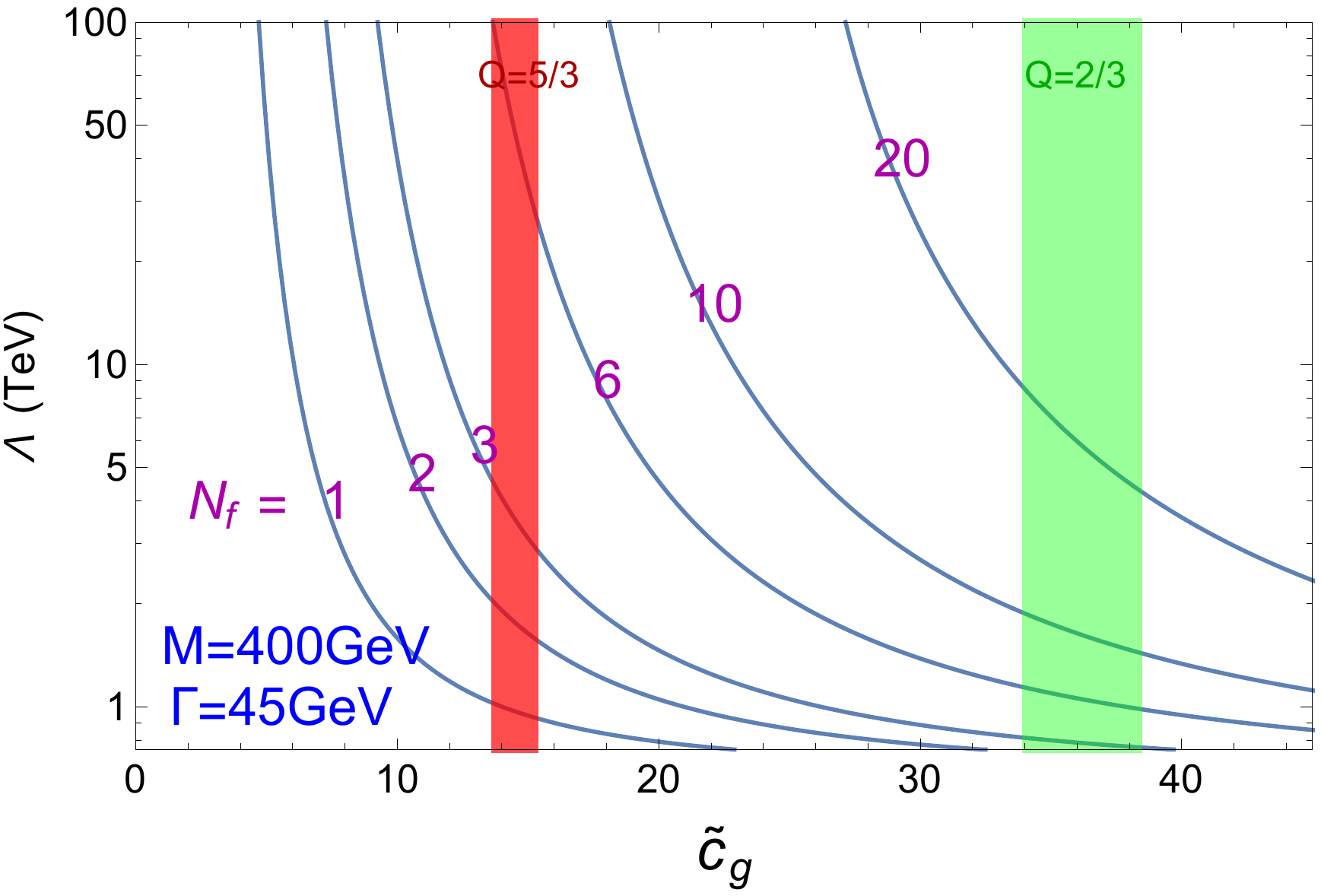}
\caption{Same as Fig.~\ref{fig:rge1} but for $S$ being a pesudoscalar. The constraints on $\tilde{c}_g$ from the dijet searches are $\tilde{c}_g < 10.1,~25.5,~66.0$ for $\Gamma = \Gamma_{\rm min}, ~1\,{\rm GeV},  ~45\,{\rm GeV}$, respectively.}
\label{fig:rge2}
\end{figure}

Our results for the cutoff scale $\Lambda$ as a function of $c_g$ ($\tilde{c}_g$) are shown in Fig.~\ref{fig:rge1} for $S$ being a scalar and Fig.~\ref{fig:rge2} for $S$ being a pseudoscalar, where different curves represent different assumptions on the value of $N_f$.   The cutoff scale has a strong dependence on the mass of the fermion(s), in particular when $M$ is small and the threshold effects become important.  To illustrate this, we show the results for two different values of $M$.  For the plots on the left panel, we assume $M=1\,$TeV; for the plots on the right panel, we assume $M=400\,$GeV, large enough to avoid on-shell decays of $S$ while giving a significant loop factor ($1.36$ for a scalar and $1.68$ for a pseudoscalar).  It should be noted that a $400\,$GeV charged quark with standard production and decays should have already been excluded by the 8\,TeV LHC.  Still the VLQ could be hidden if it decays in an exotic way due to BSM model construction, similar to the case of hidden scalar top quark. 
We also show the $1\sigma$ preferred ranges on $c_g$ ($\tilde{c}_g$) required to reproduce the observed diphoton excess for the singlet quarks ($B_R$,  $T_R$ and $X^{5/3}_R$ in Table~\ref{tab:coefficients}), displayed as bands with different colors.  The red, green and blue bands correspond to charge $5/3$, $2/3$ and $1/3$ quarks, respectively.  We note here that the signs of the charges do not matter for our results.  For simplicity, we only consider one type of quark (with $N_f$ copies) at a time, while in principle one could have a mixture of different quarks.  The bounds on $c_g$ ($\tilde{c}_g$) also depends crucially on the assumption of the total width of $S$.  In each figure, the plots in the top panel correspond to a minimal total width, while the plots in the middle (bottom) panel correspond to a total width of $\Gamma = 1\,$GeV (45\,GeV).

\begin{figure}[t]
\centering
\includegraphics[width=7.5cm]{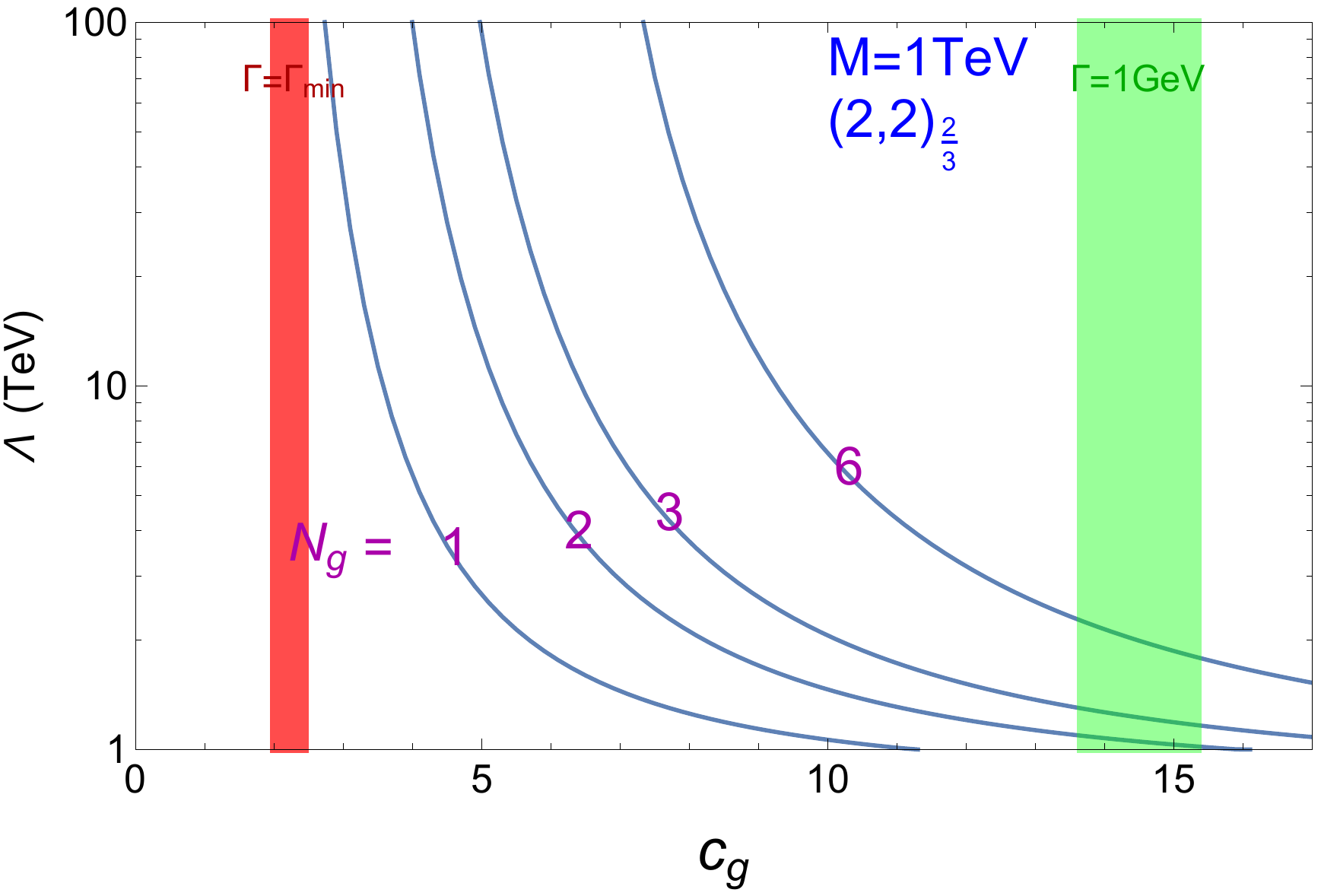}
\includegraphics[width=7.5cm]{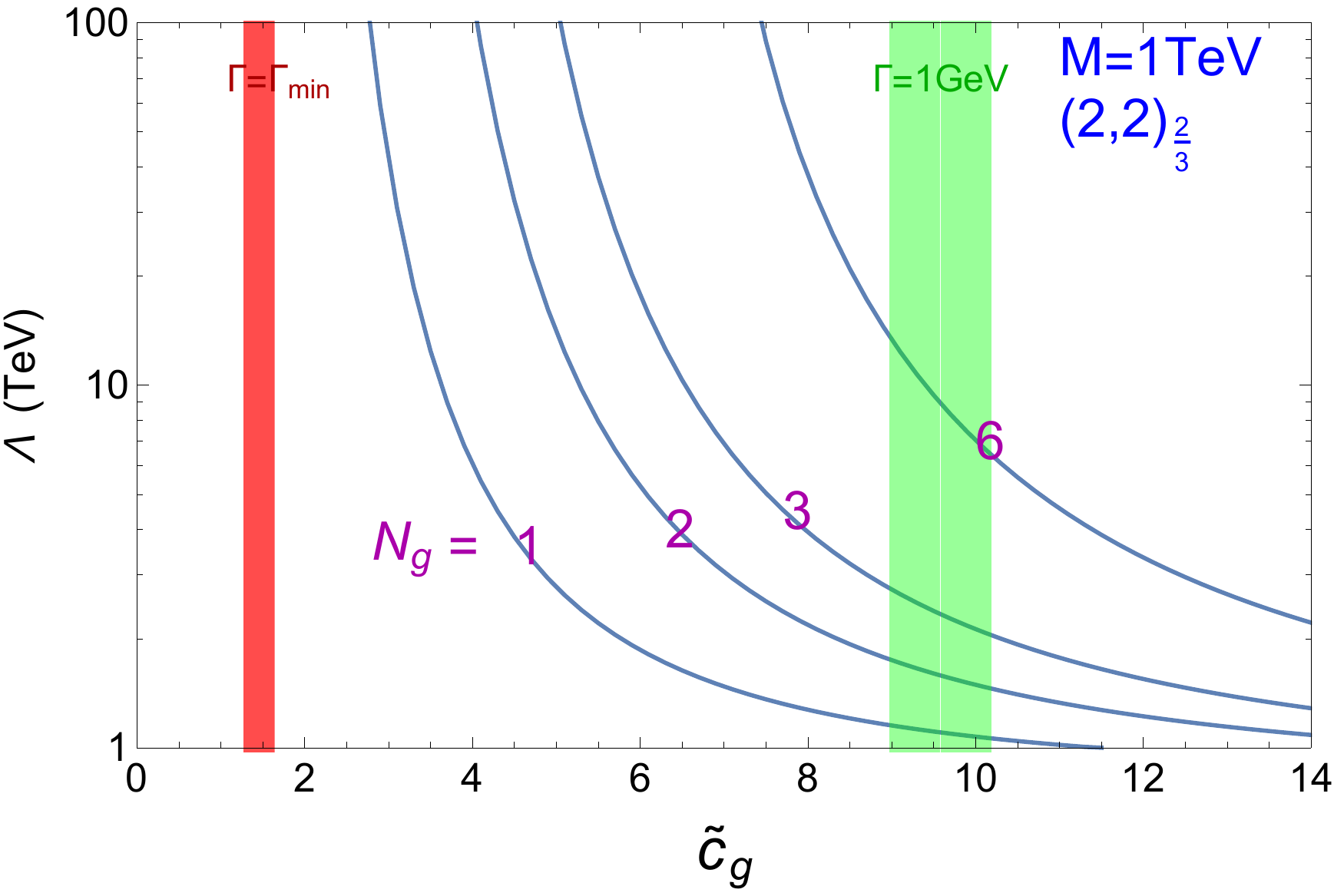}
\caption{Same as Fig.~\ref{fig:rge1} and Fig.~\ref{fig:rge2}  but for a bidoublet $(2,2)_{\frac{2}{3}}$.  $N_g$ denotes the number of copies (generations) of bidoublets.   The red, green and blue bands correspond to the $1\sigma$ preferred ranges on $c_g$ required to produce the observed diphoton excess assuming $\Gamma = \Gamma_{\rm min}$, $\Gamma =1\,$GeV and $\Gamma =45\,$GeV, respectively.  The plot on the left (right) side corresponds to $S$ being a scalar (pseudoscalar).  Note that the blue band is out of the plot ranges due to the required $c_g$($\tilde{c}_g$) being very large.  The plotted bands are also within the allowed region of $c_g$($\tilde{c}_g$) from the dijet search constraints, which are $c_g < 15.4,~38.5,~100$ (of a scalar) and $\tilde{c}_g < 10.1, ~25.5, ~66.0$ (of a pseudoscalar) for $\Gamma = \Gamma_{\rm min}, ~1\,{\rm GeV},  ~45\,{\rm GeV}$, respectively. }
\label{fig:rgex}
\end{figure}

In Fig.~\ref{fig:rgex}, we show the results for the bidoublet quark $(2,2)_{\frac{2}{3}}$, where $N_g$ denotes the number of copies (generations) of the bidoublet.  In each plot, the red, green and blue bands correspond to the $1\sigma$ preferred ranges on $c_g$ to produce the observed diphoton excess assuming $\Gamma = \Gamma_{\rm min}$, $\Gamma =1\,$GeV and $\Gamma =45\,$GeV, respectively.  The plot on the left (right) side corresponds to $S$ being a scalar (pseudoscalar).  We only show results for $M=1\,$TeV.  The masses of these quarks are already constrained to be above $\sim800$\,GeV by the 8\,TeV LHC results (see {\it e.g.}~Ref.~\cite{Chatrchyan:2013wfa}), assuming they decay dominantly to SM particles.  Since these quarks are EW doublets, it is very hard, if not impossible, to arrange them to have light masses and simultaneously avoid the LHC constraint.  We do not consider this possibility here.

A large electric charge of the quarks is very helpful to increasing the cutoff scale, as a large charge would greatly enhance $c_\gamma$ and reduce the required Yukawa coupling.  
In fact, for $\Gamma = \Gamma_{\rm min}$, having one charge $5/3$ quark could make the theory weakly coupled at all scales due to the gluon loop contribution to the running and a small initial Yukawa coupling, as discussed earlier.  On the other hand, for a charge $1/3$ quark, even with $\Gamma = \Gamma_{\rm min}$, the cutoff scale can not be much larger than a few TeV unless there are many copies of it (large $N_f$).  In addition, for charge $1/3$ VLQs the required value of $c_g$($\tilde{c}_g$) is already excluded by the dijet searches, regardless of the number of flavors.

Having a pseudoscalar instead of a scalar for the $750\,$GeV resonance could help increase the cutoff scale for two reasons.  First, for a given scenario the required $\tilde{c}_g$ is less than the required $c_g$ (of a scalar)  
due to a larger contribution from fermion loops.  Second, for small $M$ (near the threshold) the loop function of a pseudoscalar is more enhanced than the one of a scalar.   A pseudoscalar may also be convenient for other aspects, such as avoiding mixing with the SM Higgs in the CP conserving case.

Another observation here is that if the total width of $S$ is large, it is very hard to increase the cutoff scale.   
This is because a large total width makes the branching ratio to diphoton very small, and to obtain the desired rate the production cross section needs to be very large.  This large production rate requires a large $c_g$ ($\tilde{c}_g$) which in turns requires large Yukawa couplings.  
Even for $\Gamma =1\,$GeV, it is very hard to make the cutoff much larger than 1\,TeV without a combination of large electric charge, large $N_f$ and small $M$.  If the diphoton excess does come from a resonance at around $750\,$GeV, the measurement of its width is indeed crucial for the understanding of the underlying new physics.

It is important to realize that the validity of the one loop RG equations becomes questionable near the cutoff scale.  This is because of the following two reasons:  1) when the coupling becomes strong, higher order effects become important and the one loop RG equations cease to be good approximations; 2) if the underlying new physics involves some strong dynamics under which $S$ is a composite state, the theory containing $S$ as a degree of freedom is only well-defined much below the cutoff scale at which the composite states form.  Therefore, one should not interpret the results in Fig.~\ref{fig:rge1}, \ref{fig:rge2} and \ref{fig:rgex} literally as the exact value of the cutoff scale given a set of model parameters, since there is no unique and precise definition of the cutoff scale to begin with.  Instead, they should only be treated as an estimation of the scale around which one would expect to see additional new physics.

%%%%%%%%%%%%%%%%%%%%%%%%%%%%%%%%%%%%%%%%%%%%%%%%%%%%

\section{Conclusion}
\label{sec:con}

While the observed diphoton excess can be most easily explained by a (pseudo)scalar $S$ at around $750\,$GeV  with VLQ loop-induced couplings to gluons and photons, the measured rate implies relatively large Yukawa couplings between $S$ and the VLQ(s).  Such large couplings can easily run into perturbativity limit at nearby scale.  In this paper, we take a conservative approach and assume no other new physics appear except for the 750\,GeV scalar and the VLQ(s) mediating the anomalous couplings. We assess the validity of such setup in various VLQ representations in three different scalar width assumptions, namely the minimal width, a medium width of 1\,GeV and a large width of 45\,GeV. Most importantly, we estimate the highest scale that the theory is still self-consistent/perturbative, using the tools of RG running.  We explore different ``methods'' of increasing the cutoff scale and find out that introducing multiple copies ($N_f$) of quarks, quarks with light masses ($\sim400\,$GeV) or large electric charges ($5/3$) could all help.  We point out that while the initial Yukawa coupling can be significantly reduced by a large $N_f$, the running of the Yukawa will be sensitive to all flavors through the self-energy of the singlet scalar field. In addition, a more comprehensive consideration of perturbativity/unitarity requires the Yukawa coupling to satisfy $y\lesssim 4\pi / \sqrt{N_c N_f}$. These two important $N_f$ factors brings the cutoff scale sooner than one would naively expect.  The required cutoff scale also depends crucially on the total width of $S$.  We find that for the $\Gamma_{\rm min}$ case, with a $1\,$TeV charge $5/3$ VLQ with $N_f=1$, the minimal theory could be weakly coupled at all scales, as the Yukawa couplings runs down due to the gluon loop contribution when its initial value is small enough; for $\Gamma=45\,\gev$, even with $400\,$GeV charge $5/3$ VLQs, one would need at least $N_f\gtrsim 3$ to increase the cutoff scale to $\sim5\,$TeV. 

Our study is based upon very generic setups, and the conclusions are applicable for many variations of beyond standard model  physics addressing this diphoton excess.  Further studies could reveal other interesting features, for example, in considering a combination of scalar and fermion (including lepton) contributions, other production modes, and potentially a thorough study trying to discover new particles running in the loop.  
As the data from the 13\,TeV LHC accumulates, the diphoton excess is expected to be  soon confirmed or ruled out, and in the former case more measurements could be made, such as the width of the resonance and the rate of other decay modes. We look forward to the upcoming more detailed LHC run2 analysis for this exciting opportunity for new physics.

\section*{Acknowledgements}
We would like to thank Haipeng An,  Patrick Fox, Peisi Huang, Jack Kearney and Lian-Tao Wang  for helpful discussions.  Fermilab is operated by Fermi Research Alliance, LLC under Contract No. DE-AC02- 07CH11359 with the United States Department of Energy.  JG is supported in part by the Chinese Academy of Science (CAS) International Traveling Award under Grant H95120N1U7.  ZL would also like to thank the Center for Future High Energy Physics (CFHEP) in Beijing for its hospitality where part of this work was done.

%%%%%%%%%%%%%%%%%%%%%%%%%%%%%%%%%%%%%%%%%%%%%%%%%%%%%%%
%%%%%%%%%%%%%%%%%%%%%%%%%%%%%%%%%%%%%%%%%%%%%%%%%%%%%%%

\providecommand{\href}[2]{#2}\begingroup\raggedright\endgroup

%%%%%%%%%%%%%%%%%%%%%%%%%%%%%%%%%%%%%%%%%%%%%%%%%%%%%%%
%%%%%%%%%%%%%%%%%%%%%%%%%%%%%%%%%%%%%%%%%%%%%%%%%%%%%%%

\end{document}